\documentclass[acmsmall,screen]{acmart}

\usepackage{subfigure}
\usepackage{graphicx}
\usepackage{multirow}
\usepackage{booktabs}
\usepackage{lettrine}
\usepackage{hyperref}
\usepackage{makecell}

\usepackage{xcolor}
\usepackage{colortbl}  

\definecolor{lightgrey}{rgb}{0.905, 0.902, 0.902}
\usepackage{longtable}

\AtBeginDocument{%
  }

\newcounter{RubingCommentCounter}
\newcounter{DaveCommentCounter}
\newcommand{\ddu}[1]{
    \stepcounter{DaveCommentCounter}
    \textcolor{blue}{\textit{/**Dave's comment [\arabic{DaveCommentCounter}]: I don't understand the intended meaning in the next sentence. Please revise/delete/explain. **/}}
}

\newcommand{\dns}[1]{
    \stepcounter{DaveCommentCounter}
    \textcolor{blue}{\textit{/**Dave's comment [\arabic{DaveCommentCounter}]: I'm not sure that I have captured the intended meaning in the next sentence. Please check/confirm. **/}}
}

\newcounter{ZhenzhenCommentCounter}
\setcopyright{acmlicensed}
\copyrightyear{2018}
\acmYear{2018}
\acmDOI{XXXXXXX.XXXXXXX}

\acmJournal{JACM}
\acmVolume{37}
\acmNumber{4}
\acmArticle{111}
\acmMonth{8}

\begin{document}

\title{Requirements-Based Test Generation: A Comprehensive Survey}


\author{Zhenzhen Yang}
\email{3240002362@student.must.edu.mo}
\orcid{0009-0009-9358-7675}
\affiliation{
  \institution{School of Computer Science and Engineering, Macau University of Science and Technology}
  \city{Taipa}
  \state{Macau}
  \country{China}
  \postcode{999078}
}
\affiliation{
  \institution{School of Modern Information Technology, Zhejiang Polytechnic University of Mechanical and Electrical Engineering}
  \city{Hangzhou}
  \state{Zhejiang}
  \country{China}
  \postcode{310053}
}

\author{Rubing Huang}
\email{rbhuang@must.edu.mo}
\orcid{0000-0002-1769-6126}
\affiliation{
  \institution{School of Computer Science and Engineering, Macau University of Science and Technology}
  \city{Taipa}
  \state{Macau}
  \country{China}
  \postcode{999078}
}
\affiliation{
  \institution{Macau University of Science and Technology Zhuhai MUST Science and Technology Research Institute}
  \city{Zhuhai}
  \state{Guangdong Province}
  \country{China}
  \postcode{519099}
}

\author{Chenhui Cui}
\email{3230002105@student.must.edu.mo}
\orcid{0009-0004-8746-316X}
\affiliation{
  \institution{School of Computer Science and Engineering, Macau University of Science and Technology}
  \city{Taipa}
  \state{Macau}
  \country{China}
  \postcode{999078}
}

\author{Nan Niu}
\email{nan.niu@uc.edu}
\orcid{0000-0001-5566-2368}
\affiliation{
  \institution{Department of Electrical Engineering and Computer Science, University of Cincinnati}
  \city{Cincinnati}
  \state{OH}
  \country{USA}
  \postcode{210030}
}

\author{Dave Towey}
\email{dave.towey@nottingham.edu.cn}
\orcid{0000-0003-0877-4353}
\affiliation{
  \institution{School of Computer Science, University of Nottingham Ningbo China}
  \city{Ningbo}
  \state{Zhejiang}
  \country{China}
  \postcode{999078}
}


\renewcommand{\shortauthors}{Yang et al.}

\begin{abstract}
    As an important way of assuring software quality, software testing generates and executes test cases to identify software failures.
    Many strategies have been proposed to guide test-case generation, such as source-code-based approaches and methods based on bug reports. 
    \textit{Requirements-based test generation} (RBTG) constructs test cases based on specified requirements, aligning with user needs and expectations, without requiring access to the source code.
    Since its introduction in 1994, there have been many contributions to the development of RBTG, including various approaches, implementations, tools, assessment and evaluation methods, and applications.
    This paper provides a comprehensive survey on RBTG, categorizing requirement types, classifying approaches, investigating types of test cases, summarizing available tools, and analyzing experimental evaluations.
    This paper also summarizes the domains and industrial applications of RBTG, and discusses some open research challenges and potential future work.
\end{abstract}





\received{4 May 2025}

\maketitle

\section{Introduction}

Software development is a complex process driven by rapidly evolving technologies and continuously changing requirements.
It is often influenced by shifts in business processes. 
A key part of the software-development process is software testing, which helps to ensure the quality of the software.
Testing involves verifying that the software satisfies the specified requirements, and can account for over half of the total development costs \cite{mustafa2021automated}. 
There are three main stages in software testing: 
test case generation, execution, and evaluation. 
Among these, test-case generation may represent the most intellectually demanding process. 
Manual test-case generation can be time-consuming and prone to human error.

Automated test-case generation has been widely regarded as an effective approach to reducing effort and cost \cite{jin2021generation}. 
Test cases can be derived from requirements specifications, design artifacts \cite{zakeriyan2021towards}, source code \cite{hamberger2023specification}, or historical test artifacts (such as bug reports) \cite{plein2024automatic}. 
While code-based test-case generation can effectively identify errors at the implementation level, it may not address issues arising from misunderstandings or mistakes in the analysis or design phases \cite{yazdani2019automatic}. 
Faulty code can also mislead the test-generation process. 
Automating test-case generation from bug reports involves reproducing reported issues following the steps provided by users \cite{plein2024automatic, fazzini2018automatically}. 
This can be valuable for managing and analyzing a large volume of bug reports, thus reducing the manual effort required. 
However, there may be challenges, due to the informal nature of user-reported bugs, and that these reports may be mainly from the software maintenance phase. 
\textit{Requirements-based test generation} (RBTG), which relies on requirements specifications or models (before any code is written), may provide a more robust solution. 
By generating test cases early in the development lifecycle, RBTG can detect inconsistencies and ambiguities in requirements, reducing the risk of propagating errors into later stages of development. 
This early intervention can significantly lower development costs by addressing defects earlier.

Due to its simplicity, and no specialized training being needed to understand it, \textit{natural language} (NL) has often been used for software requirements' documentation \cite{dwarakanath2012litmus}. 
However, NL can be imprecise and ambiguous, potentially making it unsuitable for automated test-case generation. 
Over the years, significant research has been dedicated to addressing the challenge of RBTG \cite{rodrigues2024systematic}.  
To address the limitations of NL, semi-formal and formal representations have also been proposed \cite{fischbach2019automated}.  
RBTG can be traced back to Peng Lu in 1994 \cite{lu1994test}.
\citet{mustafa2021automated}
presented a review of RBTG from 2000 to 2018, reporting that most approaches used \textit{Unified Modeling Language} (UML) for RBTG, with a primary focus on functional testing. 
\citet{farooq2022requirement}, \citet{jin2021generation}, and \citet{ahmad2019model} also examined UML-based methodologies for test case generation. 
Recent studies have also explored the application of \textit{natural language processing} (NLP) techniques to the testing of NL-based requirements \cite{boukhlif2024natural} \cite{garousi2020nlp}. 
\citet{wang2024software} provided a comprehensive analysis of \textit{large language models} (LLMs) in software testing, while \citet{fontes2023integration} investigated the integration of \textit{machine learning} (ML) techniques for automated test generation. 
These studies focused on the application of NLP, LLMs, and ML in test case generation. 
Nevertheless, to the best of our knowledge, there is a lack of an up-to-date, comprehensive survey encompassing the entire RBTG process and potential future research directions. 
This paper aims to address this gap.

This paper presents a comprehensive survey of RBTG, covering 267 papers published up to 2024. 
It examines the entire RBTG process, including: 
(1) a summary and analysis of the selected 267 papers; 
(2) a classification of the RBTG input requirements; 
(3) a description, classification, and synthesis of RBTG approaches; 
(4) a description of the types of test cases produced; 
(5) a summary of RBTG tools; 
(6) an analysis of RBTG evaluation methods; 
(7) an overview of the RBTG application domains; and 
(8) a discussion of open research challenges and potential future directions for RBTG. 
To the best of our knowledge, this is the first large-scale and comprehensive survey on RBTG.

The rest of this paper is structured as follows. 
Section \ref{section:Background and Related Work} discusses the background and previous surveys. 
Section \ref{section:Research Methodology} explains the methodology adopted for our literature review. 
Section \ref{section:RQ1} explores the distribution of RBTG studies. 
Section \ref{section:RQ2} examines the types of RBTG input requirements. 
Section \ref{section:RQ3} reviews the approaches used for RBTG. 
Section \ref{section:RQ4} categorizes the types of test cases generated. 
Section \ref{section:RQ5} presents the tools developed for RBTG. 
Section \ref{section:RQ6} provides a detailed analysis of RBTG evaluation methods. 
Section \ref{section:RQ7} examines the application domains covered in RBTG studies. 
Section \ref{section:RQ8} identifies key RBTG challenges, and proposes future research directions. 
Finally, Section \ref{section:Conclusion} concludes the paper. 

\section{Background and Related Work
\label{section:Background and Related Work}}

This section introduces the definition and classification of requirements, and explains their importance for defining system objectives. 
We also outline the types of tests, explaining how they help to ensure that the system performs as intended. 
Previous survey studies on test-case generation are also reviewed, summarizing their key outcomes and methodologies.

\subsection{Requirements
\label{subsection:bak-requirements}}

Software requirements serve as the blueprint for the development of a software application, presenting the core goals and objectives that guide the development team \cite{umar2024advances}. 
The requirements specify the intended features and functionality, covering aspects such as product perspective, user characteristics, operating environment, design constraints, user documentation, assumptions, and dependencies \cite{bruel2021role}. 
The quality of the requirements can significantly impact the overall quality of both the software testing and development \cite{mustafa2021automated}.

In typical industrial environments, users may communicate the needs of a software application by stating the requirements in NL, using things like requirements specifications and user stories. 
Pure NL is easy to understand, making it accessible to both developers and testers. 
However, requirements described in NL may contain ambiguities or inconsistencies, and may be incomplete. 
This will impact on the quality of software testing. 
To minimize such problems, requirements can be converted into more structured and formal representations, enhancing clarity and consistency.  

\citet{fischbach2019automated} noted that requirements could be specified into three different classes: informal, semi-formal, and formal techniques. 
Informal descriptions do not follow a particular syntax or semantic structure;
semi-formal requirements are expressed in a predefined and fixed syntax that is not formally defined; and 
formal notations are based on mathematical notation, and have precise syntax and semantics.
However, it is difficult to fully represent the diverse forms of requirements within RBTG using just this classification. 
Furthermore, these broad categories lack the precision needed to align with the trends in modern requirements engineering. 

\citet{bruel2021role} categorized requirements into five groups based on their primary specification style: 
NL, semi-formal, automata/graphs, mathematical, and seamless (programming language-based).  
Semi-formal specifications structure the expression of requirements by establishing a defined language that is more precise than NL, yet does not rely on mathematical formalisms or programming-language constructs.
Examples of this kind of requirement include:
Reqtify \cite{dassault2016Reqtify}, KAOS \cite{van2001goal}, and \textit{systems modeling language} (SysML) \cite{chen2023research}.
Automata/graphs-based requirements are modeled through state machines \cite{shaheen2024case}, transition systems \cite{cartaxo2007test}, or graph-theoretical representations, such as UML \textit{activity diagrams} (ADs) \cite{kamonsantiroj2019memorization} and state diagrams \cite{chevalley2001automated}. 
Mathematical notation relies on formalisms based on set theory \cite{kadakolmath2022model, saiki2021tool} or universal algebra \cite{dick1993automating}. 
Seamless requirements are closely integrated with implementation artifacts through programming language constructs, enabling direct traceability to the code \cite{naumchev2019seamless}.   
In the context of an RBTG review study, the generation of test cases is independent of source code:
The ``seamless'' category, therefore, is problematic, as studies involving source code are typically excluded during the literature search.
Nevertheless, within the field of test-case generation methods, model-based approaches are extensively employed, with a wide variety of model types. 
The ``automata/graphs'' category outlined in this paper does not comprehensively cover all model types, nor does it account for scenarios involving the combined use of multiple requirements' types. 
To address these limitations, this paper proposes a revised classification framework that builds on the existing categories while aligning more closely with the characteristics of RBTG methods.

\begin{table}[]
\caption{Requirements Classification for RBTG}
\scriptsize
\renewcommand{\arraystretch}{1.3}
\label{tab:tableofrequirementsclassification}
    \begin{tabular}{lp{22em}p{22em}}
    \hline
    \textbf{Category}  & \textbf{Core Characteristics} & \textbf{Typical Formats}  \\
    \hline
    Natural language & Unstructured or simplified textual descriptions in human language. & Unstructured natural language \cite{aoyama2021executable}, use case textual description \cite{wang2020automatic}, user stories \cite{fischbach2020specmate} \\
    Semi-formal  & Combines natural language with structured elements such as templates, tables, or simplified graphical representations & Ontology-based requirements specification  \cite{banerjee2021ontology, ul2019ontology}, SysML  \cite{zhu2021model, chen2023research} \\
    Model-based  & Abstract representations of software behavior or functionality, using graphical, automata, or trees.  & UML diagrams \cite{linzhang2004generating, sahoo2021test}, Automata \cite{hopcroft2001introduction, linz2022introduction}  \\
    Formal mathematical& Mathematically rigorous and precise, using formal notations for specification. & \textit{Structured Object-Oriented Formal Language} (SOFL) \cite{saiki2021tool, cajica2021automatic}, Z formal specification \cite{kadakolmath2022model, helke1997automating}  \\
    Hybrid & Combine requirements from multiple categories for flexibility and comprehensiveness. & Textual specifications combined with a use case diagram, textual use cases, HTML screen mockups, and glossaries \cite{clerissi2017towards}\\
    \hline
    \end{tabular}
\end{table}

Table \ref{tab:tableofrequirementsclassification} groups the requirements representations in the surveyed papers into five categories:
\begin{itemize}
  \item 
  \textit{Natural language requirements}: 
  Expressed in English or other natural languages, these requirements are intuitive and easy to understand. 
  To improve clarity and accessibility for both technical and non-technical audiences, their expressiveness is sometimes restricted to a controlled subset of the language.
  
  \item 
  \textit{Semi-formal requirements}: 
  Combining NL with some formal structure, such as tables in text documents, or diagrams from modeling tools. 
  These approaches provide enhanced clarity and accessibility for users with minimal technical training, but cannot be directly processed by formal verification methods. 
  
  \item 
  \textit{Model-based requirements}: 
  Representing software behavior or functionality through abstract representations, such as graphical models, automata, or trees. 
  These representations improve understanding and communication for both technical and non-technical users by using visual aids without necessitating deep knowledge of the underlying mathematics. 
  These models (e.g., UML-based models) have a long tradition in software engineering \cite{uzun2018model}. 
  
  \item 
  \textit{Formal mathematical requirements}: 
  Beyond those of the previous categories, these requirements are specified using rigorous mathematical notations (such as those based on set theory or algebra), allowing precise definitions suitable for formal verification. 
  However, these are typically only accessible to technical audiences with a strong foundation in mathematics, making them less appropriate for non-technical stakeholders.
  
  \item 
  \textit{Hybrid requirements}: 
  Combining requirements from two or more categories, hybrid requirements combine the strengths of each to provide a more comprehensive and detailed representation. 
  This approach ensures both breadth and accuracy in capturing requirements information, making it adaptable to diverse project needs.
\end{itemize}

\subsection{Types of Tests
\label{subsection:Types of tests}}

There are different types of tests, such as \textit{abstract test cases} (ATCs), \textit{concrete test cases} (CTCs), and \textit{test scenarios} (TSs). 
A test case specifies a particular condition to be evaluated, and includes a set of test inputs, execution conditions, 
and expected results developed to verify a specific aspect of system behavior. 
A test case may consist of a title, precondition, test steps, and expected result. 
The definitions used in this paper conform to standard terminology \cite{board2014standard}.
\begin{itemize}
  \item 
  \textit{Abstract test cases} are high-level test cases without concrete (implementation level) values for input data and expected results. 
  Logical operators are used; 
  instances of the actual values are not yet defined and/or available (e.g., given an abstract value ``$n$ is a positive integer''). 
  It is typically in a human-readable description. 

  \item 
  \textit{Concrete test cases} are low-level test cases with concrete (implementation level) values for input data and expected results. 
  Logical operators from high-level test cases are replaced by actual values that correspond to the objectives of the logical operators (e.g., a concrete value for $n$ is $100$).

  \item 
  \textit{Test scenarios} are documents specifying a sequence of actions for the execution of a test. 
  These consist of test-design specifications, test-case specifications, and/or test-procedure specifications (semi-formal specifications that lie between descriptive system requirements and executable test cases). 
  Test scenarios have been treated as ATCs in several studies \cite{sabbaghi2021framework}.
\end{itemize}

In software testing, CTCs can either be executed manually (for example, by test engineers) or automatically, using testing frameworks such as the Selenium framework \cite{waitchasarn2023generating} or JUnit \cite{uzun2018model}. 
However, some of the reviewed studies did not actually generate test cases:
This paper considered those test artifacts lacking concrete input values or expected results as ATCs.

\subsection{Previous Survey Studies}

Previous work has explored the challenge of generating test cases directly from software requirements, addressing the problem using a wide range of methods and technologies. 
This paper provides a comprehensive review of these efforts, synthesizing the existing body of knowledge and categorizing the diverse strategies employed in RBTG. 
Table \ref{tab:tableforrelatedwork} provides a comparison of prior survey studies against this paper.

\begin{table}[!t]
\caption{Comparison of Previous Survey Studies}
\label{tab:tableforrelatedwork}
\scriptsize
\renewcommand{\arraystretch}{1.3}
    \begin{tabular}{lp{16em}ccp{22em}}
    \hline
    \textbf{Year} & \textbf{Studies} & \textbf{\# Studies} & \textbf{Time Span} & \textbf{Description}  \\
    \hline
    2011 & An overview on test generation from functional requirements \cite{escalona2011overview} & 24  & 1988-2009 & This survey examined approaches for describing functional requirements informally and formalized the comparative study through characterization schema. \\
    2017 & A comprehensive investigation of natural language processing techniques and tools to generate automated test cases \cite{ahsan2017comprehensive} & 16  & 2005–2016 & This survey explored NLP techniques and tools for automating test case generation from initial requirements documents. \\
    2018 & Automated Regression Test Case Generation for Web Application: A Survey \cite{gupta2018automated} & 79  & 1988-2017 & This survey focused on the generation of regression test cases and the approaches to web applications. \\
    2019 & Model-based testing using UML activity diagrams: A systematic mapping study \cite{ahmad2019model} & 41  & 2004-2016 & This survey provided a comprehensive overview of existing approaches to model-based testing using UML ADs. \\
    2021 & Automated Test Case Generation from Requirements: A Systematic Literature Review \cite{mustafa2021automated} & 29  & 2000-2018 & This survey focused on functional testing, examining approaches and input requirement types for generating test cases. \\
    2021 & Generation of Test Cases from UML Diagrams - A Systematic Literature Review \cite{jin2021generation} & 62  & 1999-2019 & This survey investigated the generation of test cases from UML diagrams, focusing on key aspects such as model types, intermediate formats, case studies, coverage criteria, execution methods, and test case optimization.  \\
    2022 & Requirement-Based Automated Test Case Generation: Systematic Literature Review \cite{farooq2022requirement}  & 30  & 2010-2021 & This survey evaluated techniques for automated test case generation using requirements as input. It primarily focused on UML diagrams and identified the challenges and gaps.  \\
    2023 & Current Trends in Automated Test Case Generation \cite{potuzak2023current}  & 67  & 2000-2022 & This survey addressed automated test data generation and tests based on it, mapping and categorizing existing methods while summarizing their common features. \\
    2023 & Software Test Case Generation Tools and Techniques: A Review \cite{verma2023software}  & 125 & 2011-2021 & This survey conducted a systematic study of testing tools and techniques without being limited to those based on requirements. \\
    2024 & Natural Language Processing-Based Software Testing: A Systematic Literature Review \cite{boukhlif2024natural} & 24  & 2013-2023 & This survey provided an in-depth analysis of NLP-based software testing, exploring NLP techniques, software testing types, key challenges, and insights into the application of NLP in software testing. \\
    2024 & Software Testing With Large Language Models: Survey, Landscape, and Vision \cite{wang2024software} & 104 & 2020-2023 & This paper presented a detailed discussion of software testing tasks where LLMs are frequently used, with test case preparation and program repair being the most prominent. \\\hline
    2025 & This survey  & 267 & 1994-2024 & This survey investigated the entire process of RBTG, including input representations, generation approaches, supporting tools, evaluation methods, and application domains. It also highlighted open research challenges and proposed potential future directions.\\
    \hline
    \end{tabular}
\end{table}

\citet{boukhlif2024natural} presented an overview of diverse NLP techniques used to address various linguistic challenges. 
They provided an analysis of the distribution of software testing contexts across the reviewed research papers, noting a wide range of testing methodologies and objectives. 
They noted the critical importance of both machine- and deep-learning techniques in NLP for enhancing the capabilities and performance of language-centric applications. 
However, their study did not specify which techniques were best suited for particular testing methodologies. 
Nor did they explain how these integrated NLP and machine/deep learning approaches were applied to advance software testing practices.

\citet{mustafa2021automated} focused on functional testing, presenting input requirements and the approaches used for requirements transformation. 
However, they did not establish a clear connection between the requirements types and the corresponding methods, leaving a gap in how these elements are aligned. 
Similarly, \citet{farooq2022requirement} explored the generation of functional test cases, with particular attention on the use of UML diagrams:
They identified several gaps and challenges in RBTG, offering insights into areas requiring further investigation.

\citet{ahmad2019model} provided a comprehensive overview of model-based testing using UML ADs. 
They note that only a limited number of their surveyed approaches have been validated against realistic, industrial case studies, with most focusing on very restricted application domains. 
They also report that UML ADs are not commonly employed for non-functional testing, and that there is a lack of holistic or comprehensive approaches for model-based testing using UML ADs. 

Previous reviews primarily focused on input requirements and detailing the approaches and tools utilized in RBTG. 
However, these studies lack a comprehensive analysis of the interconnections between requirements, methods, tools, and processes within the context of the complete RBTG workflow.
To address this gap, this paper reports on a \textit{systematic literature review} (SLR): 
It seeks to overcome limitations of previous work, and to provide deeper insights into the different types of requirements, and their corresponding methodologies, tools, and evaluation frameworks.
The paper examines the complete RBTG workflow.

\section{Research Methodology
\label{section:Research Methodology}}

We conducted a comprehensive investigation of RBTG using the SLR methodology proposed by \citet{kitchenham2009systematic}. 
As part of the SLR, we adopted the snowballing technique, which starts with an initial set of papers and expands by examining their references to identify additional relevant studies \cite{wohlin2014guidelines}. 
The research methodology consisted of several key steps: 
defining research questions; 
identifying relevant studies; 
applying the snowballing technique to expand the search;
implementing exclusion criteria; and 
systematically extracting the required data. 
This structured approach ensures a rigorous and comprehensive analysis of the literature. 

\subsection{Research Questions}

\begin{figure}[!t]
  \centering
  \includegraphics[width=\linewidth]{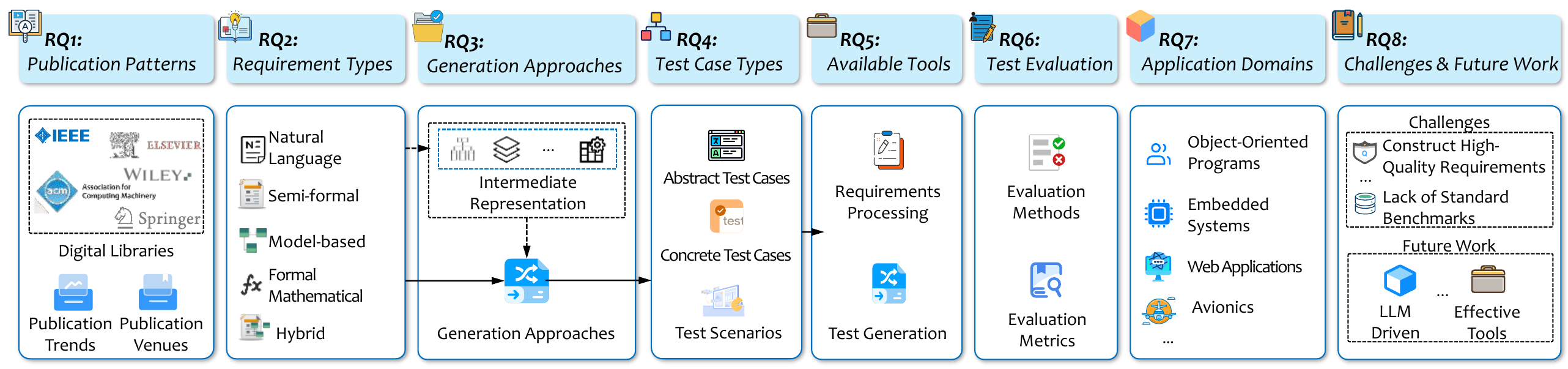}
  \caption{Requirements-based test-generation process.}
  \Description{Requirements-based test-generation process.}
  \label{fig:survey}
\end{figure}

The research objectives for this survey were based on the complete RBTG process (Figure \ref{fig:survey}), and involved analyzing and categorizing information at each stage. 
The study also aimed to identify research challenges and areas requiring future work. 
To achieve these objectives, we defined the following research questions (RQs):
\begin{itemize}
    \item 
    \textit{RQ1: What are the publication patterns of RBTG studies?}     
    
    \item 
    \textit{RQ2: What types of requirements have been used in RBTG?}
    
    \item 
    \textit{RQ3: What are the approaches that support RBTG?}
    
    \item 
    \textit{RQ4: What types of test-case representations have been explored in RBTG?}
    
    \item 
    \textit{RQ5: What tools have been developed to support RBTG?}
    
    \item 
    \textit{RQ6: How have RBTG approaches been evaluated?}
    
    \item 
    \textit{RQ7: In what domains and industrial applications has RBTG been applied?}
    
    \item 
    \textit{RQ8: What are the remaining challenges and future work for RBTG?}
\end{itemize}

\subsection{Literature Search and Selection}

The search strategy was optimized to identify relevant information, ensuring a thorough and effective investigation of the research questions. 
This process involved four steps: 
selecting appropriate digital libraries; 
identifying additional search sources;
determining relevant search strings; and 
performing the snowballing technique. 
Following previous review studies \cite{ahmad2019model, jin2021generation, mustafa2021automated, verma2023software}, we selected five online source repositories to perform the SLR: 
\begin{itemize}
  \item ACM Digital Library
  \item Elsevier Science Direct
  \item IEEE Xplore Digital Library
  \item Springer Online Library
  \item Wiley Online Library
\end{itemize}
These selected repositories provide a wide range of scholarly resources highly relevant to the research topic. 

After identifying the relevant literature databases, a search string was applied to each. 
The search strings were developed iteratively, being refined based on preliminary results. 
For each repository, customized search strings were constructed to optimize the advanced search process. 
The overall strategy for constructing the strings was: 
(1) The \textit{\textbf{Title}} should include \textit{test case OR test suite OR test scenario}, 
(2) The \textit{\textbf{Title}} should also include \textit{generate OR create OR requirement}, 
(3) The \textit{\textbf{Abstract}} should include \textit{requirement OR specification OR user stories}. 
These three criteria were combined using the \textit{AND} operator to construct the search strings.

The search was conducted in January 2025, capturing all studies published before that date. 
Table \ref{tab:tableforsearch} shows the initial search results: 
These were then de-duplicated and screened based on titles and abstracts, as reported in the third column. 
To further enhance our literature search, we employed 
Snowballing techniques \cite{wohlin2014guidelines} were used to enhance the literature search:
The references of selected papers were reviewed, and additional relevant studies were identified (and downloaded from Google Scholar). 
This iterative process was continued until no new relevant papers were identified. 
The final results are listed in the ``Post-Snowballing'' column. 
The final set of relevant papers was determined by applying the following exclusion criteria:
\begin{enumerate}
  \item Studies without a full text.
  \item Studies not written in English.
  \item Books, book chapters, or technical reports (most of which may have been published as articles).
  \item Masters or Ph.D. theses.
  \item Survey studies.
  \item Studies less than five pages in length.
\end{enumerate}

\begin{table}[!t]
	\caption{Overview of Search Result and Study Selection}
    \footnotesize
	\label{tab:tableforsearch}
	\begin{tabular}{lcccc}
		\hline
		\textbf{Source} & \textbf{\begin{tabular}[c]{@{}c@{}}Initial Search\\ Results\end{tabular}} & \textbf{\begin{tabular}[c]{@{}c@{}}After De-duplication \&\\ Title/Abstract Review\end{tabular}} & \textbf{Post-Snowballing} & \textbf{\begin{tabular}[c]{@{}c@{}}Final After\\ Exclusions\end{tabular}} \\
		\hline
		ACM  & 76  & 30  & 49  & 31 \\
		ELSEVIER & 46  & 12  & 19  & 14 \\
		IEEE & 314 & 127 & 182 & 113 \\
        Springer & 34  & 6  & 57  & 51 \\
		Wiley  & 23  & 4 & 6 & 3  \\
		Other  & -- & -- & 55  & 55 \\\hline
		\textbf{\textit{Total}}  & 493 & 179 & 368 & 267 \\  
		\hline  
	\end{tabular}
\end{table}

Table \ref{tab:tableforsearch} also shows the distribution of selected studies across the various digital libraries: 
45\% from IEEE Xplore; 
18\% from SpringerLink; 
12\% from ACM Digital Library; 
5\% from Elsevier; 
1\% from Wiley; and 
19\% from other databases (such as ProQuest, Academia.edu, and SCIRP.org), identified through snowballing techniques. 
This diverse distribution highlights the collaborative and extensive nature of academic research in the field.

Although it may not have been possible to find every relevant paper through our search, we are confident that we have included most of the significant publications. 
This survey provides a comprehensive overview of the current trends and state-of-the-art in RBTG.

\subsection{Data Extraction}

The 267 selected studies were thoroughly reviewed and inspected, with data extracted according to the research questions:
Table \ref{tab:tablefordata} summarizes the extracted data. 
To minimize the risk of missing information and reduce potential errors, we performed reviews both during the data-extraction process, and after its completion.

\begin{table}[]
\caption{Data Collection for Research Questions}
\footnotesize
\label{tab:tablefordata}
\centering
    \begin{tabular}{ll}
    \hline
    \textbf{RQs} & \textbf{Type of Data Extracted} \\ \hline
    RQ1  & Core information for each paper (publication year, publisher, type of paper).  \\ 
    RQ2  & Type of requirements, descriptions, classifications. \\ 
    RQ3  & Type of techniques, degree of automation, intermediate representations, descriptions. \\ 
    RQ4  & Type of generated output artifacts and test execution environment. \\ 
    RQ5  & Tool name, function, language, types, and descriptions. \\ 
    RQ6  & Evaluation methods, metrics. \\ 
    RQ7  & Application domains, programs in case studies.  \\ 
    RQ8  & Details of remaining challenges, potential future work.  \\ \hline
    \end{tabular}
\end{table}

\section{Answer to RQ1: What are the Publication Patterns of RBTG Studies?
\label{section:RQ1}}

267 studies were identified for critical evaluation regarding RBTG, with the extracted data summarized in Table \ref{tab:tablefordata}. 
Core information was extracted for each study, including publication year, publisher, and type of paper. 
This section provides an overview of the surveyed studies, focusing on their publication trends and venues.

\subsection{Publication Trends}

\begin{figure}[!t]
  \centering
  \subfigure[Number of publication per year.]{
  \label{fig:trend1}
  \includegraphics[width=0.48\linewidth]{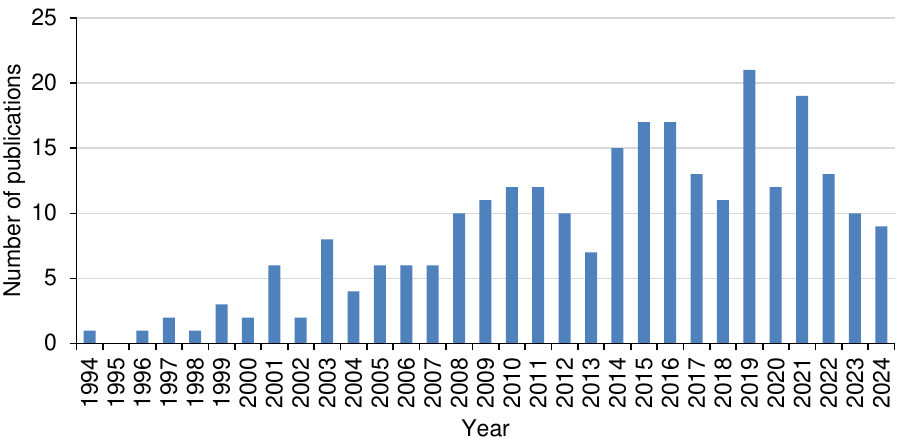}}
  \subfigure[Cumulative number of publications per year.]{
  \label{fig:trend2}
  \includegraphics[width=0.48\linewidth]{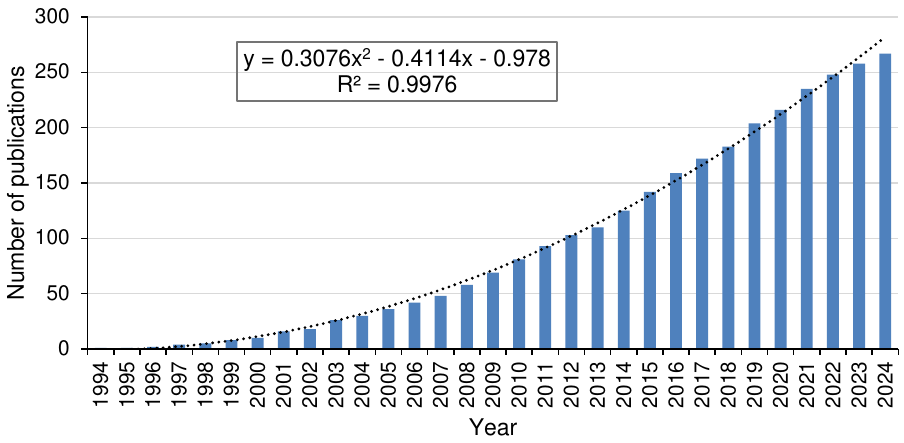}}
  \caption{Overview of the surveyed 267 RBTG studies' publication.}
  \Description{Overview of the surveyed 267 RBTG studies' publication up to 2024.}
  \label{fig:trend}
\end{figure}

Figure \ref{fig:trend} provides a comprehensive overview of the yearly RBTG publication trends of the 267 primary studies, up to the end of 2024:
Figure \ref{fig:trend1} shows the number of publications each year; and 
Figure \ref{fig:trend2} presents the cumulative number of publications over time. 
The earliest reviewed publication was from 1994. 
This field has attracted significant attention and interest from researchers since 2001. 
Since 2008, there have been at least ten published studies each year, with notable peaks of over 20 papers in 2019 and 2021. 
Figure \ref{fig:trend2} shows the polynomial trendline, with a high determination coefficient ($R^{2}=0.9976$): 
There is both a sustained and growing interest in RBTG, highlighting its importance in software-testing automation.

\subsection{Distribution of Publication Venues}

\begin{figure}[!t]
  \centering
  \subfigure[Overall venue distribution.]{
  \label{fig:venue1}
  \includegraphics[width=0.34\linewidth]{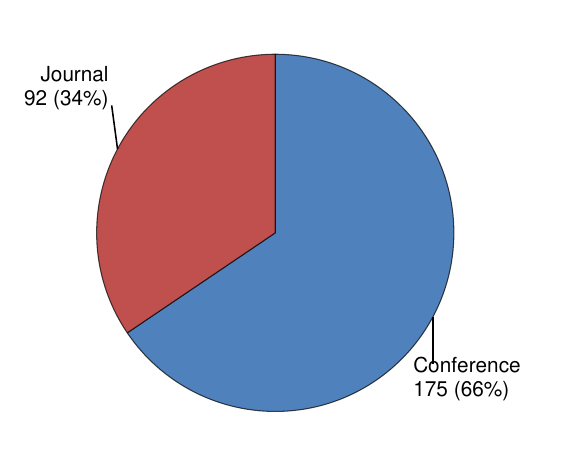}}
  \subfigure[Venue distribution per year.]{
  \label{fig:venue2}
  \includegraphics[width=0.60\linewidth]{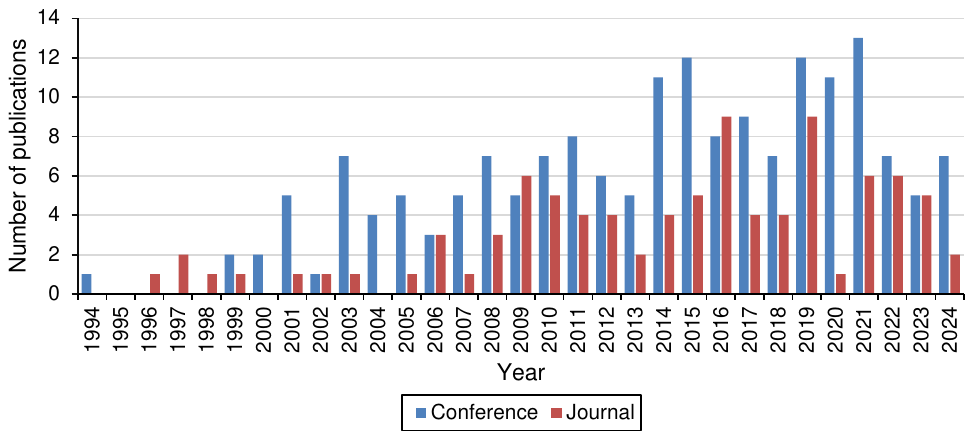}}
  \caption{Venue distribution for RBTG studies.}
  \Description{.}
  \label{fig:venue}
\end{figure}

Figure \ref{fig:venue} shows the distribution of publications across conferences and journals, with Figure \ref{fig:venue1} presenting the overall venue distribution, and Figure \ref{fig:venue2} illustrating the distribution per year. 
As shown in Figure \ref{fig:venue1}, 175 (66\%) of the 267 primary studies were published in conferences and 92 (23\%) were in journals. 
The number of studies published in conferences has consistently exceeded those in journals over the years, reflecting conferences as key platforms for disseminating new research findings quickly and effectively.

\section{Answer to RQ2: What Types of Requirements Have Been Used in RBTG?
\label{section:RQ2}}

Requirements play a crucial role in RBTG, as their content and representations directly influence generation strategies and approaches, ultimately affecting the quality of the resulting test cases. 
In real-world projects, requirements are often written in NL, which is easy to understand, but inherently ambiguous. 
To address this, NL requirements are often transformed into formal or semi-formal specifications. 
This section categorizes the collected papers based on requirement types, identifying the most commonly used representations, and analyzing their evolution over time.

\begin{figure}[!b]
  \centering
  \subfigure[Overall requirements types.]{
  \label{fig:requirement1}
  \includegraphics[width=0.41\linewidth]{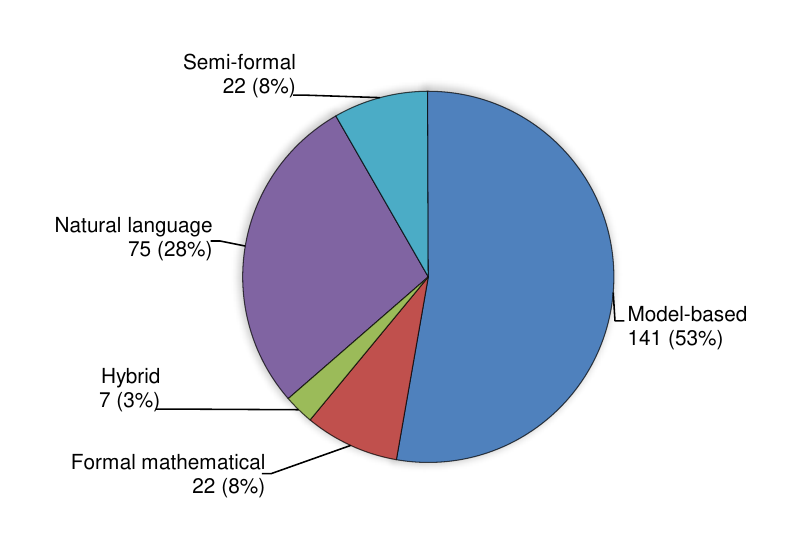}}
  \subfigure[Requirements types per year.]{
  \label{fig:requirement2}
  \includegraphics[width=0.55\linewidth]{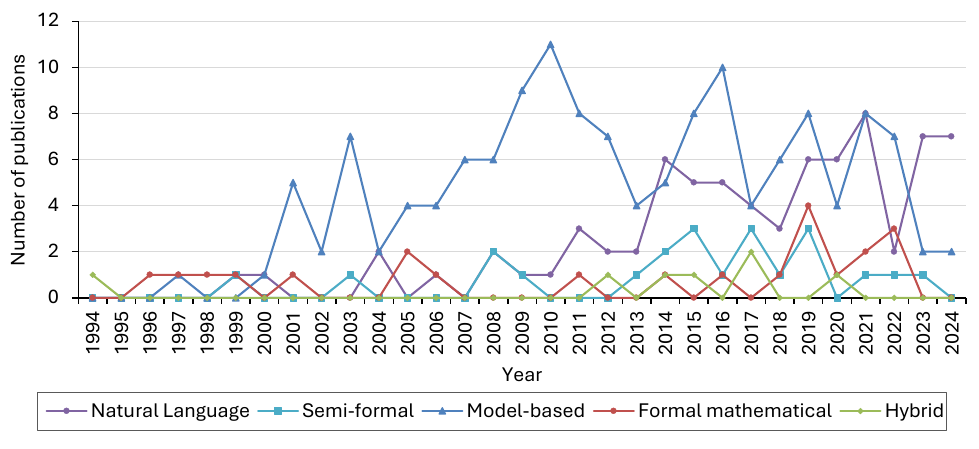}}
  \caption{Types of requirements.}
  \Description{Types of requirement.}
  \label{fig:requirements}
\end{figure}

As explained in Section \ref{subsection:bak-requirements}, this paper uses five requirements categories: 
NL;
semi-formal;
model-based;
formal mathematical; and 
hybrid. 
Figure \ref{fig:requirement1} presents the distribution of the types of requirements, showing that 
53\% (141 papers) focused on model-based representations; 
28\% (75 papers) involved NL; and 
19\% (51 papers) used other representations. 
Figure \ref{fig:requirement2} shows the requirements' types over time:
Although model-based requirements have been extensively studied over the years, NL types began to attract attention in 2011, and have surpassed model-based approaches since 2019.

\subsection{Natural Language Requirements}

Users usually communicate the needs of a software application in NL. 
This has meant that generating test cases directly from NL requirements has become a key objective for researchers. 
Among the surveyed studies, the NL-based requirements can be categorized into three types: 
unstructured;
templated; and 
\textit{controlled natural language} (CNL) requirements. 
Unstructured NL refers to free-text expressions with no constraints on the requirements' formulation:
This is the most commonly used approach in practice. 
Templated requirements include use case descriptions, user stories (which are commonly employed in agile development), and scenarios. 
A use case is a sequence of interactions between a system and a user (understood as an instance of an actor) \cite{ryser1999scenario}. 
A scenario depicts a portion of the application's behavior at a specific time and place, capturing the situational context \cite{sarmiento2014c}:
Use cases can be seen as distinct models of scenarios. 
CNL requirements use restricted grammar and predefined vocabulary to reduce textual complexity and ambiguity.  

\begin{figure}[!t]
  \centering
  \includegraphics[width=0.6\linewidth]{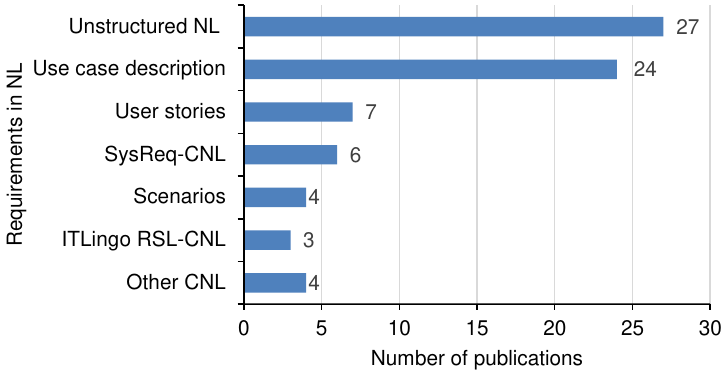}
  \caption{Distribution of types of natural language requirements.}
  \Description{Distribution of types of natural language requirements.}
  \label{fig:requirements-nl}
\end{figure}

Figure \ref{fig:requirements-nl} presents the distribution of different NL types used in the surveyed studies. 
Unstructured natural language \cite{masuda2021syntax, arora2024generating} represents the main form of input requirements, used by 27 studies (36\%). 
The use case description \cite{mahalakshmi2018named, sneed2018requirement} is also widely adopted, appearing in 24 studies (32\%). 
Other notable NL types include user stories (seven papers), scenarios (four papers), 
CNL like SysReq (six papers), and ITLingo RSL (three papers). 
SysReq is a CNL for system requirements, defined by a phrase-structure context-free grammar and a lexicon encompassing the application domain vocabulary \cite{carvalho2014nat2testscr, silva2015test}.
ITLingo RSL (Requirements Specification Language) is a controlled and integrated natural language, designed to assist in the systematic, rigorous, and consistent production of requirements specifications \cite{paiva2019requirements, miranda2020preliminary}.
Although NLP and 
ML techniques have also been combined to directly transform these requirements into test cases \cite{arora2024generating, ueda2024automatic}, the majority of studies use an intermediary step, converting the requirements into structured models before generating the test cases. 
Further details about these approaches will be explored in Section \ref{section:RQ3}.  

\subsection{Semi-formal Requirements}

Semi-formal requirements combine NL with structured templates, tables, or simplified graphical representations.
Examples include ontology-based specifications \cite{banerjee2021ontology}, SysML \cite{chen2023research}, and courteous logic representation \cite{sharma2014natural}.  

Within this category, the requirements ontology (five papers) and SysML (three papers) were the most frequently used. 
The requirements ontology represents concepts and relationships within a domain \cite{banerjee2021ontology, ul2019ontology}, with inference rules describing strategies for deriving test cases from the ontology \cite{tarasov2017application}. 
SysML models the \textit{system under test} (SUT) along with its interactive external environment \cite{zhu2021model}, providing diagrams that enable the textual depiction of requirements, including non-functional ones \cite{chen2023research, abbors2009tracing}. 
Semantic specifications \cite{moitra2019automating, crapo2019using} that are formulated in the controlled English \textit{semantic application design language} (SADL) map directly into the \textit{web ontology language} (OWL), ensuring readability and comprehension for both humans and machines. 
Courteous logic-based requirements representations \cite{sharma2015natural, Sharma2014automated}, \textit{expressive decision tables} (EDTs) \cite{venkatesh2015cost, venkatesh2015generating} and \textit{domain specific language} (DSL) \cite{cabral2008requirement, olajubu2017automated} have also been adopted, each appearing in two studies. 
Despite their potential to streamline test generation, only 22 studies (8\%) used semi-formal requirements: 
These were mostly applied to embedded systems (six papers) and safety-critical domains, like avionics and automotive (seven papers).

\subsection{Model-based Requirements
\label{subsection:Requirements In Model-based types}}

\begin{figure}[!t]
  \centering
  \includegraphics[width=\linewidth]{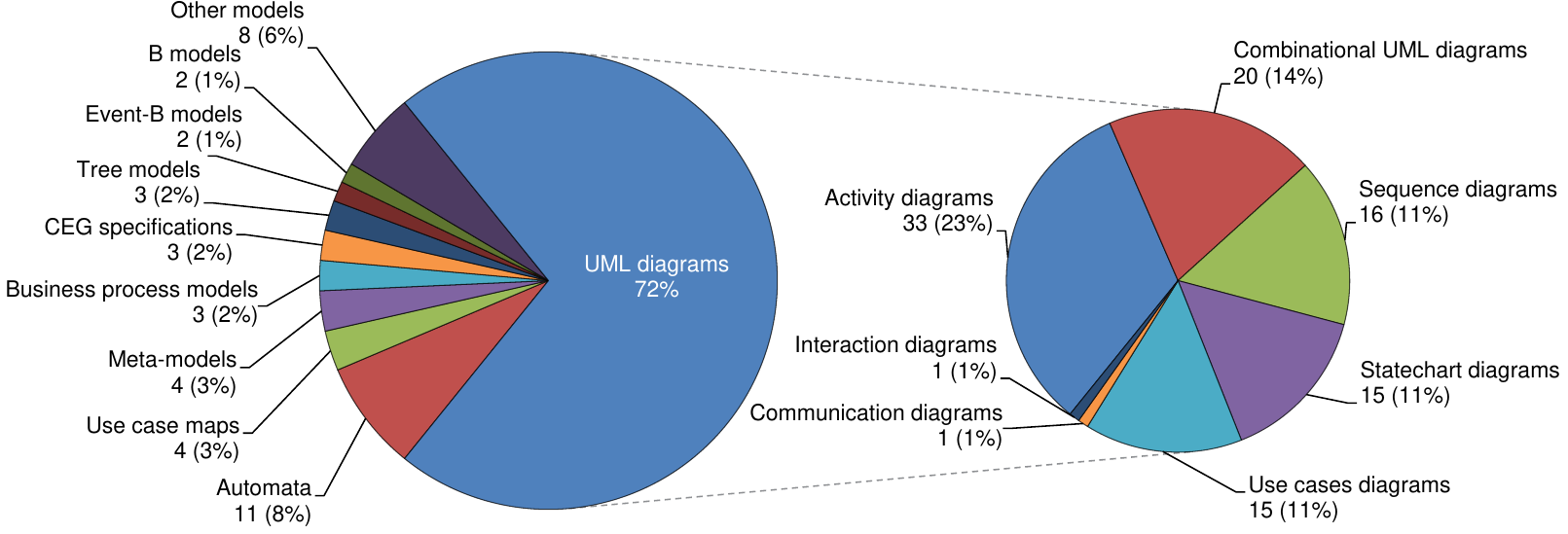}
  \caption{Distribution of model-based types.}
  \Description{Distribution of model-based types.}
  \label{fig:requirements-models}
\end{figure}

Requirements models provide an intuitive and visual representation of system requirements. 
This section examines the models used in the surveyed studies. 
Figure \ref{fig:requirements-models} presents the distribution of requirements models across the studies. 
The majority of papers (72\%) used UML diagrams, highlighting the significant focus on UML in the literature. 
Automata models, such as \textit{finite state machines} (FSMs) and \textit{extended finite state machines} (EFSMs), were used in 8\% of the papers. 
Other models included: 
use case maps (3\%), meta-models (3\%), business process models (2\%), specifications (2\%), tree models (2\%), B models (1\%), and event-B models (1\%).
There were also several other uncommon models, each appearing only once. 

\subsubsection{UML Diagrams}

The UML has been widely used as a system-level specification language. 
It is a visual modeling language that comprises various diagrams, each representing different aspects of a system's structure and behavior. 
Figure \ref{fig:requirements-models} shows that UML ADs \cite{linzhang2004generating, mingsong2006automatic, fan2009test} (23\%) are the most commonly used model:
As a behavioral model, ADs provide detailed implementation information, making them well-suited for test derivation. 
However, their effectiveness decreases when dealing with large and complex systems.
Combinational UML diagrams (14\%), sequence diagrams (11\%), statechart diagrams (11\%) and use case diagrams (11\%) have also been widely used. 
Combinational UML diagrams combine the strengths of multiple models, enabling the generation of more reliable test cases
---
however, they also incur a time overhead \cite{perez2018automatic}. 
Examples of combinational UML diagrams include activity and statechart diagrams \cite{sahoo2021test, swain2010test2}, sequence and use case diagrams \cite{sarma2007automatic, swain2010test} and sequence and statechart diagrams \cite{khurana2015test, sokenou2006generating}. 
Both sequence \cite{khandai2011novel, samuel2008test} and statechart diagrams \cite{hametner2011test, pradhan2022transition} are also behavioral models, capturing system interactions and state transitions to support test generation. 
In contrast, use case diagrams \cite{hamza2021analyzing, nebut2006automatic} describe system functionality from a user's perspective, providing a high-level view of interactions without detailing execution flow or state changes.
Communication diagrams \cite{samuel2007automatic} convey both the messages exchanged between objects and their execution order, providing insight into both data flow and control flow. 
Similarly, interaction diagrams \cite{nayak2009model} visually depict the sequence of messages exchanged among collaborating objects across different scenarios of a use case.

\subsubsection{Automata}

Automata \cite{hopcroft2001introduction, linz2022introduction} are abstract mathematical machines used to model the behavior of systems, usually through states and transitions between them. 
Four types of automata were used in the surveyed papers: 
EFSM \cite{elqortobi2023granular, rao2016search};
FSM \cite{rechtberger2022prioritized, liu2010new};
safety SysML state machine (S2MSM) \cite{wang2022mc};
and Mealy machines \cite{pfaller2008requirements}. 
EFSM was the most commonly used, appearing in five papers, followed by FSM, which appeared in four. 
Both S2MSM and Mealy machines were found in one paper each. 
Table \ref{tab:tableofautomata} compares the different automata models. 

\begin{table}[!t]
  \caption{Automata Models}
  \scriptsize
  \label{tab:tableofautomata}
  \begin{tabular}{p{12em}llp{20em}}
  \hline
  \textbf{Automata Model} & \textbf{Tuple Structure} & \textbf{Mathematical Definition}  & \textbf{Description}  \\
  \midrule
  \textit{Finite state machine} (FSM) & 5-tuple & $M = (S, s_0, I, O, T )$  & \begin{tabular}[c]{@{}l@{}}$S$ is nonempty finite set of states,\\ $s_0 \in S$ is the initial state,\\ $I$ and $O$ denote nonempty finite sets of inputs or\\outputs respectively,\\ $T$ is the finite set of transitions\end{tabular} \\
  \textit{Extended finite state machine} (EFSM) & 6-tuple & $M = (S, s_0, V, I, O, T )$  & $V$ is the finite set of (internal) context variables \\
  \textit{Safety sysML state machine} (S2MSM) & 7-tuple & $M = (S, T, E, G, P, A, V)$ & \begin{tabular}[c]{@{}l@{}}$E$ is a finite set of events,\\ $G$ is a finite set of guards,\\ $P$ is a finite set of the priority,\\ $A$ is a finite set of actions,\\ $V$ is a finite set of variables\end{tabular}  \\
  \textit{Mealy machine} & 6-tuple & $M = (S, s_0, I, O, \delta, \lambda)$  & $\delta : S \times I \rightarrow S$ and $\lambda : S \times I \rightarrow O$ are total functions which denote the deterministic transition and output relations \\
  \hline
  \end{tabular}
\end{table}

An FSM is a finite labeled transition system where each transition is associated with an input/output pair, consisting of an input action and a corresponding output action \cite{ibias2021coverage, liu2010new}. 
An EFSM system model is an extension of FSMs that includes additional features, like variables and conditions on transitions \cite{chen2018automatic, vaysburg2002dependence}. 
S2MSM has additional verification and refinement procedures to check the safety of the generated model using a formal verification tool:
S2MSM can both precisely model functional requirements and capture safety concerns, in a unified model \cite{wang2022mc}. 
A Mealy machine is an FSM where the outputs are determined by both the current state and the current input. 
It enables compact modeling of reactive systems by associating outputs directly with transitions, allowing for precise representation of input–output behaviors \cite{pfaller2008requirements}.
Due to their logical structure, these automata are well-suited for modeling the structural and behavioral aspects of software applications, particularly in testing. 
However, their effectiveness is limited when handling more complex data dependencies and complex systems.

\subsubsection{Use Case Map}
A use case map (UCM) is a visual notation for modeling scenarios/workflows derived from requirements \cite{kesserwan2019use}. 
It is similar to a UML AD, but imposes fewer constraints, emphasizing the causal relationships between workflow steps without specifying detailed message exchanges or data \cite{boucher2017transforming}. 
Four of the surveyed papers used UCMs as input \cite{boucher2017transforming, kesserwan2019use, boulet2015towards, amyot2005ucm}.

\subsubsection{Meta-Model}
Meta-models standardize the information used in test-case generation: 
They formally specify a concept by defining its attributes and its relationships with other concepts \cite{gutierrez2012automatic}.
\citet{gutierrez2012automatic, gutierrez2015model} used a functional requirements meta-model as the basis for testing: 
The core concept of this meta-model was the functional requirements' element, defined by 
(1) \textit{step} elements representing discrete actions (e.g., data requests or computations); and 
(2) execution-order elements specifying their sequence. 
\citet{chen2022improved} proposed a test-requirement meta-model that defined the core concepts along with their relationships and dependencies. 
\citet{granda2014towards} applied transformation rules to convert meta-models into test cases.

\subsubsection{Other Types of Models}
In addition to the models already discussed, several others were also used. 
\textit{Business process models} (BPMs) \cite{yazdani2019automatic, khader2016utilizing, gupta2011model} represent a series of related activities producing a specific service or product. 
A \textit{Cause-effect graph} (CEG) \cite{krupalija2022forward, kalaee2016optimal, paradkar1997specification} is a black-box testing technique that describes relationships between events using logical relations (and events are labeled as either causes or effects). 
A \textit{Behavior tree} (BT) model \cite{lindsay2015automation, salem2010requirement} is a directed tree with various node types and branching, aiding in modeling complex behaviors. 
Additional models include:
Event-B models \cite{malik2009scenario, satpathy2006synthesis}; 
B models \cite{bouquet2005requirements, legeard2001generation}; 
scenario-based state/event trees \cite{tsai2003scenario}; 
EDTs \cite{agrawal2020scaling};
\textit{variable relation models} (VRMs) \cite{wang2020automatic, zhipeng2021method}; and 
\textit{multiple condition-control flow graphs} (MCCFGs) \cite{son2019mccfg}. 
These models all provide different approaches for representing requirements.

\subsection{Formal Mathematical Requirements}

Formal mathematical specifications use precise logical formulas and constraint-specification languages to express system requirements, typically using mathematical notations \cite{bruel2021role}.  
A \textit{Structured object-oriented formal language} (SOFL) \cite{saiki2021tool, cajica2021automatic, sato2015automatic} integrates graphical and formal notations, defining functions through pre- and post-conditions. 
Z formal specification \cite{kadakolmath2022model, helke1997automating, chang1998testing, stocks1996framework} structures specifications into modular units called schemas using formal mathematical notation. 
\textit{Linear temporal logic} (LTL) \cite{aniculaesei2019using, aniculaesei2019automated, whalen2006coverage} describes the temporal evolution of system behaviors with mathematical formulas:
It has been widely used in formal verification, especially for model checking. 
Other formal specifications, such as \textit{software cost reduction} (SCR) specifications \cite{gargantini1999using}, \textit{functional scenario form} (FSF) \cite{liu2020automatic} and \textit{featured transition systems} (FTSs) \cite{devroey2014abstract}, have also been used.
Although these specifications offer advantages (like automatic generation of test cases), they also face disadvantages in terms of meeting coverage requirements efficiently.

\subsection{Hybrid Requirements}

The hybrid category includes requirements specifications that combine two or more types of representations. 
A combined textual specification, for example, may consist of a use case diagram, textual use cases, HTML screen mockups, and a glossary defining key terminology \cite{clerissi2017towards}. 
Other combinations have also been observed, such as UML diagrams integrated with business vocabulary and business rules \cite{essebaa2019model} or object constraint language \cite{ali2014test}, as well as visual requirement specifications or prototypes \cite{singi2015model}.
Among the surveyed papers in this review, seven (3\%) used hybrid requirements. 
Future requirements may likely integrate multiple types of formal specifications \cite{clerissi2017towards}: 
This hybrid approach helps to ensure the correctness and completeness of requirements, supporting the generation of effective test cases.  

\section{Answer to RQ3: What Are the Approaches that Support RBTG?
\label{section:RQ3}}

\begin{figure}[!t]
  \centering
  \includegraphics[width=0.7\linewidth]{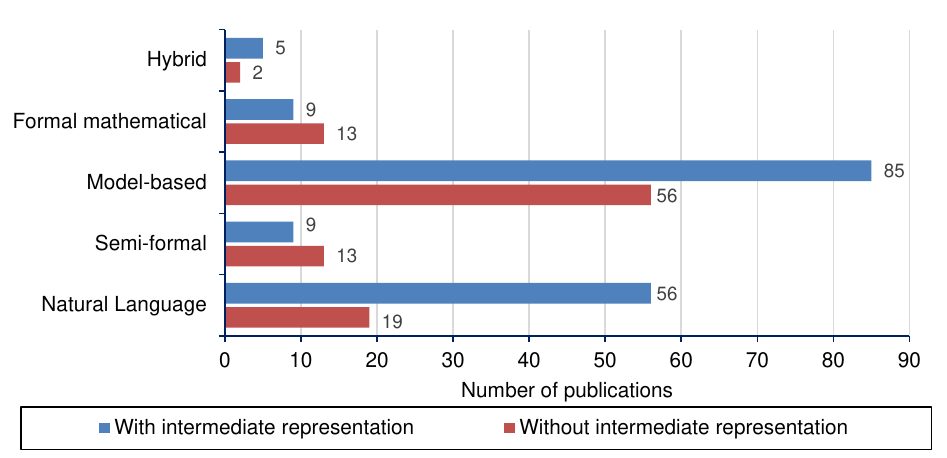}
  \caption{Test-case generation approaches with/without intermediate representations.}
  \Description{Test-case generation approaches with/without intermediate representations.}
  \label{fig:approaches}
\end{figure}

RBTG generation approaches take requirements as input and produce test cases as output. 
Different requirements' representations use different strategies to generate test cases. 
Of the 267 surveyed studies, 39\% generated test cases directly from the requirements, with the other 61\% first transforming the requirements into intermediate representations (before generating test cases). 
Figure \ref{fig:approaches} shows that most common forms of requirements (such as models and NL) use intermediate representations as the basis for test-case generation.
Accordingly, this section examines these two main strategies: 
direct test-case generation and test-case generation through intermediate representations.

\subsection{Test-Case Generation without Intermediate Representations}

\begin{table}[!b]
\caption{Approaches without Intermediate Representations}
\footnotesize
\label{tab:tableofwithoutintermediate}
    \begin{tabular}{llc}
    \hline
    \textbf{Requirements} & \textbf{Approach} & \textbf{No. of Studies}  \\
    \hline
    \multirow{5}{*}{Natural language} & \textit{Natural language processing} (NLP)  & 7  \\
      & \textit{Machine learning} (ML)  & 2  \\
      & Model-based approach & 2  \\
      & Hybrid approach  & 2  \\
      & Other specific approach & 6  \\\hline
    \multirow{5}{*}{Semi-formal}  & Ontology‑based approach  & 4  \\
      & Model-based approach & 3  \\
      & Semantic-head-driven approach  & 2  \\
      & ASSERT™  & 2  \\
      & \textit{Row-guided random algorithm with fuzzing} (RGRaF) & 2  \\\hline
    \multirow{6}{*}{Model-based} & Model-based approach & 31 \\
      & Heuristics-based approach  & 5  \\
      & Transformation-based approach  & 3  \\
      & Hybrid approach  & 2  \\
      & Other specific approach & 15 \\\hline
    \multirow{3}{*}{Formal mathematical} & Model-based approach & 6  \\
      & Heuristic-based approach  & 3  \\
      & Specific approach & 5  \\\hline
    Hybrid  & Other specific approach & 2  \\
    \hline 
    \end{tabular}
\end{table}

Table \ref{tab:tableofwithoutintermediate} summarizes the direct generation approaches identified in the surveyed papers. 
These approaches are categorized based on the types of input requirements:
The form of the requirements naturally shapes the generation strategy. 
Within each category, the approaches are arranged in the order of their frequency of occurrence in the reviewed studies. 

In the context of ``Natural Language'' requirements, NLP remains the most widely adopted:
This includes techniques such as syntax testing \cite{intana2020syntest}, which analyzes and captures the sequence of patterns or syntax of variables from the data dictionary; 
syntactic similarity \cite{masuda2016syntactic}, which determines the case conditions and actions; and
keyword extraction \cite{ansari2017constructing}. 
Recently, LLMs have gained significant attention and are increasingly becoming a preferred approach for test-case generation \cite{arora2024generating, mathur2023automated, yin2024leveraging, shakthi2024automated}. 
ML has also been commonly used: 
Ueda et al. \cite{ueda2021accuracy}, for example, enhanced the accuracy of automatic test-case generation by training data selection. 
Similarly, Kikuma et al. \cite{kikuma2018automatic} trainind ML on tagged data, and then applied the model to new requirements specifications to generate test cases. 
Model-based approaches have also been used to derive test cases from CNL requirements, making use of the characteristic CNL constructs. 
However, these approaches may require substantial manual effort \cite{paiva2020requirements, maciel2019requirements}. 
Hybrid approaches integrate multiple methodologies: 
Fernandes et al. \cite{fernandes2020towards}, for example, integrated equivalence class partitioning and boundary value analysis, while Elghondakly et al. \cite{elghondakly2015waterfall} combined text mining with symbolic execution methodologies. 
Specific tailored approaches have also been proposed:
Dwarakanath et al. \cite{dwarakanath2012litmus}, for example, introduced Litmus, which uses a syntactic parser (called the Link Grammar parser) alongside pattern-matching rules. 
These rules were designed to evaluate testability, identify test intents, and systematically generate both positive and negative test cases
---
a positive test case verifies the system's expected behavior under normal conditions, and a negative test case checks the system's response to exceptional or invalid inputs.  

``Semi-formal'' approaches primarily depend on the structure and form of the input requirements. 
These approaches include those based on ontologies, such as the approaches by \citet{ul2019ontology} and \citet{feldmann2014keeping}, which integrate a knowledge-based system (ontology) with a learning-based testing algorithm to automate the test-case generation. 
Similarly, \citet{tarasov2017application} used inference rules to query the ontology, enabling the verification of conditions and the retrieval of the necessary data for test-case construction. 
Model-based approaches are also included in this category, where test cases are derived from SysML models \cite{zhu2021model, abbors2009tracing}.
The semantic-head-driven generation approach \cite{sharma2015natural}, produced functional test cases using knowledge stored in a courteous logic \cite{lafi2021automated} representation of requirements
--- 
this is an expressive subclass of standard logical representations that incorporates procedural attachments for prioritized conflict resolution. 
Hybrid approaches have also been proposed, such as the $ASSERT^{TM}$ \cite{moitra2019automating, crapo2019using, li2019requirements}:
This approach uses a suite of tools designed to capture requirements;
provides explainable and automated formal analyses to detect and resolve errors during the requirements authoring process; and 
automatically generates a complete set of test cases. 
The RGRaF \cite{venkatesh2015cost, venkatesh2015generating} applied a dependency-driven random algorithm with fuzzing to generate test cases, targeting time-boundary conditions.

In the ``Model-based'' category, naturally, model-based approaches represent the main strategy, with the core methodology centered on constructing detailed models derived from requirements. 
Test cases are extracted using traversal algorithms, such as \textit{depth first search} (DFS) \cite{linzhang2004generating}, pre-order traversal algorithms \cite{yonathan2024generating}, and category partition methods \cite{gutierrez2015model}. 
Several heuristics-based approaches have also been applied for model-based requirements: 
Genetic algorithms \cite{ibias2021coverage, shirole2010hybrid},
ant-colony algorithms \cite{agrawal2020scaling}, and 
metaheuristic techniques \cite{khalifa2019efficient} have all been used to generate test cases. 
A genetic algorithm uses iterative stochastic optimization through selection, crossover, and mutation to evolve test-case populations, converting objectives into quantifiable fitness functions, and evaluating generated test data. 
An ant colony algorithm, inspired by the way ants use pheromones to find routes between colonies and food, can solve computational problems and identify optimal paths within graphs. 
Metaheuristics, as problem-independent optimization strategies, are black-box techniques capable of finding global optima without relying on greedy search methods. 
Transformation-Based Approaches \cite{boucher2017transforming, hue2018transformation} represent specific implementations of model-based strategies, focusing on unique traversal mechanisms. 
Hybrid approaches combine multiple methods, such as equivalence and \textit{classification tree methods} (CCTM) \cite{intana2019impact}, to improve testing efficiency and coverage.

Model-based and heuristic-based techniques are common in ``formal mathematical'' approaches. 
\citet{kadakolmath2022model} introduced a ProZ model-checking-based approach that uses a breadth-first search strategy to generate test cases. 
\citet{aniculaesei2019using} used the NuSMV model checker to derive test cases from LTL requirements. 
\citet{liu2020automatic} proposed a vibration method to heuristically generate test cases, and \citet{wang2019specification} used a genetic algorithm to identify the best mutant test condition, which was then used to generate test data.

Among the approaches that generate test cases directly from requirements (without intermediate models), model-based approaches are very versatile and applicable across multiple requirements' forms. 
They are also the most widely adopted strategy. 
In addition to model-based techniques, heuristic-based approaches and NLP are also frequently used. 
Among the heuristic-based approaches, genetic algorithms are particularly prominent. 
Recently, NLP has become an increasing popular technique for test-case generation.

\subsection{Test Case Generation with Intermediate Representations}

Many test-case generation approaches use an intermediate representation as part of the process:
This helps with the extraction and formalization of key information from requirements, improving both the accuracy and effectiveness. 
As Figure \ref{fig:approaches} shows, 61\% of the surveyed studies use an intermediate model as part of the transformation process. 

\subsubsection{Intermediate Representations}

This section explores the different types of intermediate representations commonly used in the reviewed studies.

\begin{table}[!b]
\caption{Intermediate Representations}
\centering
\footnotesize
\label{tab:tableofintermediaterepresnetations}
    \begin{tabular}{llcc}
    \hline
    \textbf{\begin{tabular}[c]{@{}l@{}}Intermediate Representation\\ 
    Categories\end{tabular}} & \textbf{Intermediate Representations} & \textbf{No. of Studies} & \textbf{Total} \\
    \hline 
    \multirow{10}{*}{Graphical models} & \textit{Activity diagram} (AD) or \textit{activity graph} (AG)  & 19 & \multirow{10}{*}{90} \\
     & Combined directed graphs  & 10 &  \\
     & \textit{Control flow graph} (CCFG) & 9  &  \\
     & \textit{Sequence diagram} (SD) or \textit{sequence graph} (SG) & 8  &  \\
     & Trees & 7  &  \\
     & \textit{State shart diagram or statechart diagram graph} (SCDG) & 6  &  \\
     & \textit{State transition diagram} (STD)  & 4  &  \\
     & \textit{Cause-effect-graph} (CEG)  & 3  &  \\
     & \textit{Use case test models} (UCTMs) & 3  &  \\
     & \textit{Flow dependency graph} (FDG) & 2  &  \\
     & Other diagrams or other directed graph  & 19 &  \\\hline
    \multirow{8}{*}{State/Process models}  & \textit{Finite state machine} (FSM)  & 11 & \multirow{8}{*}{47}  \\
     & \textit{Communicating sequential processes} (CSP) test models & 6  &  \\
     & \textit{Extended finite state machine model} (EFSM)  & 4  &  \\
     & Meta-model  & 2  &  \\
     & Symbolic model  & 2  &  \\
     & Knowledge graph & 2  &  \\
     & Other specific models  & 20 &  \\\hline
    \multirow{5}{*}{Logical specifications}  & \textit{Functional scenarios form} (FSF) & 3  & \multirow{5}{*}{20}  \\
     & Requirement interfaces  & 2  &  \\
     & Requirement specification ontology  & 2  &  \\
     & \textit{Software cost reduction} (SCR) specifications  & 2  &  \\
     & Other formalization & 11 &  \\\hline
    \multirow{2}{*}{Petri nets}  & \textit{Coloured petri nets} (CPN) & 2  & \multirow{2}{*}{4} \\
     & Petri-net models  & 2  & \\
    											 \hline
    \multirow{2}{*}{Tabular formats}  & \textit{Activity dependency table} (ADT)  & 1  & \multirow{2}{*}{3} \\
     & Decision table  & 2  &  \\\hline
    \end{tabular}
\end{table}

Table \ref{tab:tableofintermediaterepresnetations} groups the intermediate formats into five categories: 
graphical models; 
state/process models;
logical specifications;
Petri nets; and 
tabular representations. 
Graphical models include diagrams and directed graphs that represent system states and transitions. 
State/process models are formal models that describe system behavior with mathematical notations.  
Logical specifications use specification-based expressions to define requirements. 
Petri nets, with their unique notation and formal semantics, are treated as a separate category. 
Finally, tabular representations arrange requirements' logic into structured tables for enhanced clarity and interpretability. 
As Table \ref{tab:tableofintermediaterepresnetations} shows, graphical models were the most frequently utilized intermediate format, followed by state/process models. 
Logical specifications, Petri nets, and tabular formats are comparatively less common.  

ADs \cite{sarmiento2014c, hamza2021analyzing, sarmiento2014automated} or activity graphs \cite{chen2023research, kamonsantiroj2019memorization, swain2013generation} are the most commonly used graphical model:
This aligns with the findings in Section \ref{subsection:Requirements In Model-based types}. 
The second most popular models were combined directed graphs (10 papers):
These integrate multiple representations, such as \textit{control-flow graphs} (CFG) and \textit{data-flow graphs} (DFG) \cite{elqortobi2023granular, el2020test}, or statechart diagram graphs and sequence-diagram graphs \cite{lakshminarayana2020automatic}. 
The third most popular were control flow graphs (nine papers) \cite{lafi2021automated, swain2010test2, bansal2013model}, which provide a static representation of models by capturing all possible control flow paths. 
Other graphical formats include sequence diagram/graph \cite{ekici2023automatic, panigrahi2018test, dhineshkumar2014approach}, trees \cite{gupta2023decision, cunning1999test, tiwari2013approach}, SCDG \cite{pradhan2022transition, frohlich2000automated}, STD \cite{chatterjee2010prolific}, CEG \cite{fischbach2023automatic}, UCTMs \cite{wang2020automatic2}, and FDG \cite{kaur2018automatic}. 
Additionally, 19 other studies used various other specialized diagrams or directed graphs tailored to specific contexts \cite{wang2022mc, tiwari2015approach, yin2017automated, khurana2016novel, li2013extenics}.

FSMs were the most widely adopted state/process models, with 11 studies exploring their use \cite{ali2021model, clerissi2017towards, bin2012functional}. 
\textit{Communicating sequential processes} (CSP) models \cite{nogueira2019test, nogueira2016automatic, nogueira2014test} were also frequently used for formalizing system behaviors. 
Other intermediate models include EFSM \cite{rocha2021model, jiang2011automation}, meta-models \cite{allala2022generating, allala2019towards}, symbolic models \cite{schwarzl2010test}, and knowledge graphs \cite{nayak2020knowledge, verma2013generation}. 
Various other specialized models were also explored \cite{fujita2024method, bhatt2022requirements}.

Other intermediate representations that were less common included
logical specifications \cite{saiki2021tool, sato2020specification, carvalho2014nat2testscr, chinnaswamy2024user, charoenreh2019enhancing, carvalho2013test}, Petri nets \cite{silva2019cpn, sarmiento2016test, reza2011model}, and tabular representations \cite{noikajana2008web}. 
However, even though less frequently used, these formats offer unique advantages, and may support specialized test-generation approaches.

In summary, the transformation of raw requirements into intermediate representations has often employed specialized algorithms. 
Diagrams and directed graphs are the most popular choice, highlighting their effectiveness in test-case generation. 
However, alternative intermediate formats (such as formal models and formal specifications) may advance test-generation methodologies and enhance their applicability.

\subsubsection{Test Case Generation from Intermediate Representations}

Methods for generating test cases based on intermediate models align closely with the ``model-based'' category in Table \ref{tab:tableofwithoutintermediate}. 
Model-based approaches \cite{khandai2011test, nayak2011synthesis} dominate this field, 
followed by heuristics-based methods \cite{shirole2011generation}.
These studies highlight that, in most research, requirements are first transformed into specific models, which then support the efficient extraction of the test cases:
Various algorithms and techniques are then applied to generate the test cases, such as traversal algorithms \cite{nogueira2014test, dalai2012test, achimugu2021improved, septian2017automated}, category-partition method \cite{swain2010test3, hartmann2000uml, hierons1997testing},  vibration method \cite{saiki2021tool, liu2011vibration},
and other specialized techniques \cite{almeida2013automatic, pechtanun2012generation, de2006generating}.

\section{Answer to RQ4:  What Types of Test-Case Representations Have Been Explored in RBTG?}
\label{section:RQ4}

Test cases can be generated using different approaches and representations, which then structure the outputs into various forms. 
This section examines on the types of test cases produced. 
Through an analysis of the reviewed papers, we identified three primary categories: 
Abstract test cases (ATCs); 
Concrete test cases (CTCs); and
Test scenarios (TSs). 
(Detailed definitions of these categories were provided in Section \ref{subsection:Types of tests}.) 

\begin{figure}[!t]
    \centering
    \subfigure[Overall distribution of test-case types.]{
    \label{fig:TestCase1}
    \includegraphics[width=0.43\linewidth]{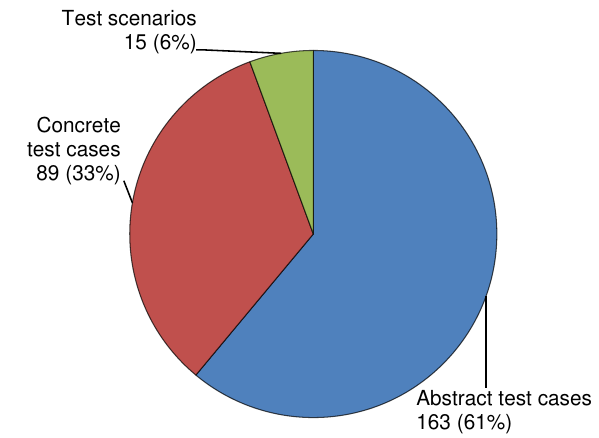}}
    \subfigure[Test-case distribution by requirement input.]{
    \label{fig:TestCase2}
    \includegraphics[width=0.54\linewidth]{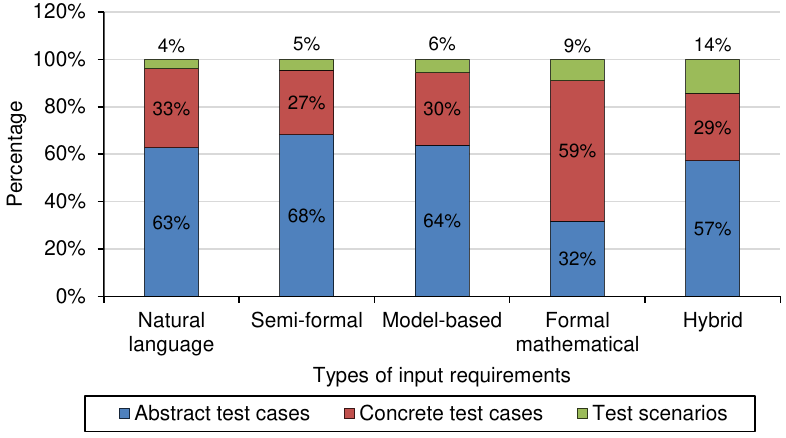}}
    \caption{Distribution of different test types.}
    \Description{Distribution of generated test cases.}
    \label{fig:testcase}
\end{figure}

The distribution of each type of test output is shown in Figure \ref{fig:TestCase1}. 
ATCs make up the majority of outputs, accounting for 61\% of the reviewed studies. 
CTCs make up 33\%, and can be executed either manually or automatically. 
Finally, TSs represent only 6\% of the test outputs.
In this paper, studies that specified the manual conversion of ATCs into CTCs were categorized as ATCs rather than CTCs:
This helps to distinguish between test outputs that are directly executable and those requiring intermediate manual processing. 
As shown in Figure \ref{fig:TestCase2}, ATCs account for more than 60\% of the generated outputs for NL, semi-formal, and model-based requirements;
ATCs represent 57\% of outputs from hybrid requirements. 
Requirements expressed through formal mathematics, in contrast, result in less than 32\% ATCs:
This is because mathematical notations and applications in fields such as signal processing or control systems often define precise input-output relationships, making them more suitable for the direct generation of CTCs, rather than ATCs.

ATCs are high-level descriptions of testing requirements that are human-readable. 
They cannot be directly executed:
They require additional modifications, such as incorporating concrete test data and defining test oracles, to transform them into executable test cases. 
The reviewed papers identified several common representation types for ATCs, including: 
\textit{textual test cases}, \textit{test paths}, \textit{tabular test cases}, \textit{test cases in other formats}.
\begin{itemize}
    \item 
    {\texttt{Textual test cases}}: 
    These closely resemble manually written test cases, and are easy for humans to understand and analyze. 
    Textual ATCs are often derived from NL or semi-formal input requirements. \citet{yin2024leveraging} used LLMs to generate textual ATCs from NL requirements. 
    \citet{fujita2024method}, \citet{ueda2021accuracy}, and \citet{fujita2024method} used ML techniques to generate textual test cases. 
    \citet{crapo2019using} and \citet{sharma2015natural} proposed approaches to obtain test cases from semi-formal requirements.
    
    \item 
    {\texttt{Test paths}}: 
    Test paths are derived by traversing specific graphs or models using specialized algorithms. 
    \citet{gupta2023decision}, for example, generated test paths from root to leaf nodes. 
    \citet{hamza2019generating}, \citet{kamonsantiroj2019memorization} and \citet{kim2007test} extracted test paths from activity graphs. 
    \citet{lafi2021automated} and \citet{alrawashed2019automated} derived test paths from control flow graphs.
    
    \item 
    {\texttt{Tabular test cases}}: 
    These use decision tables with test inputs and outputs being represented as true/false values. 
    \citet{aoyama2020test} and \citet{silva2019cpn}
    derived test cases by defining input steps based on the true or false conditions specified in the decision tables.
    \citet{almohammad2017reqcap}, \citet{olajubu2017automated}, and \citet{kalaee2016optimal} also used decision tables for test-case representation.  
    The tabular form is particularly useful for industrial control and embedded systems, where inputs and outputs are often digital signals.  
    
    \item 
    {\texttt{Test cases in other formats}}: 
    There are also several other less common representations. 
    \citet{gnesi2004formal}, for example, generated test cases in a language combining elements of process algebra that offered a formal description of system behavior. 
    \citet{chinnaswamy2024user} used regular expressions, and 
    \citet{kadakolmath2022model} constructed test cases using specified predicates, storing them in XML format.
\end{itemize}

CTCs are low-level test cases with concrete (implementation-level) values as input data and expected results. 
CTCs are more closely aligned than ATCs with actual execution. 
They can be executed (manually, or automatically using test scripts) in a specific environment.
\begin{itemize}
  \item 
  {\texttt{Manual execution}}: 
  Manually-executed CTCs often augment ATCs with detailed test data. 
  The data can be generated through manual input, randomization, or specialized methods tailored to specific applications. 
  \citet{gupta2023test} and \citet{wang2015automatic} used semi-automatic techniques to generate textual CTCs that were then manually executed. 
  \citet{sun2016transformation}, \citet{jena2014novel}, and \citet{zhang2014systematic} developed test-case generation approaches by first randomly filling values along test paths without considering input-variable constraints, and then selecting cases that met the guard conditions. 
  
  \item 
  {\texttt{Automated execution}}: 
  Automated CTC execution is primarily driven by test scripts, which can be generated using advanced tools and frameworks. 
  \citet{fernandes2020towards} used AutomTest, a tool developed in Python 3, to produce executable CTCs. 
  \citet{aoyama2021executable} and \citet{khan2022automating} converted ATCs into PROMELA test programs for their specific test execution environments. 
  \citet{granda2021towards} generated test scripts based on relations between requirements, GUI elements, and Sikulix code in the Eclipse editor. 
  \citet{maciel2019requirements}, \citet{clerissi2017towards}, and \citet{amalia2022application} used Selenium WebDriver to create executable test scripts, and \citet{pradhan2022transition} implemented test cases within the TestNG framework of J-Unit. 
\end{itemize}

TSs are documents specifying a sequence of actions for the execution of a test. \citet{arora2024generating} noted that test scenarios simulate diverse user interactions and system behaviors. 
These scenarios can be represented in various formats, such as step sequences in UML activity diagrams, sequence diagrams, or NL descriptions. 
\citet{some2008approach} generated test scenarios based on path sequences, each comprising a test-scenario identifier, a setup section, a sequence of test steps, and an expected outcome.  

In summary, test cases can be derived as ATCs in textual form or as textual test paths generated through graph traversals. 
However, these test cases often lack sufficient detail and completeness, and are unready for direct execution. 
CTCs can be generated semi-automatically with human intervention, bridging some of these gaps. 
Nevertheless, achieving a fully automated pipeline for both test-case generation and execution remains essential to significantly enhance testing efficiency and minimize manual effort. 
This transition from ATCs to CTCs poses substantial challenges and is an area of ongoing research.

\section{Answer to RQ5: What Tools Have Been Developed to Support RBTG?
\label{section:RQ5}}

Across the reviewed literature, we identified 88 tools designed specifically to support RBTG tasks, collectively referred to as RBTG tools. 
Table \ref{tab:tools} classifies these RBTG tools into two groups, according to their operational phases:  
\textit{requirements processing}, and \textit{test generation}. 
(If a tool supports both requirements processing and test-case generation, it was classified under the test generation category.) 
The table includes the tools' names (e.g., jUCMNav), relevant reference (e.g., [6]), and implementation  language (e.g., Java):
Interested readers can find the original studies by consulting the references.

\begin{table}[!t]
\caption{RBTG Tools}
\centering
\footnotesize
\label{tab:tools}
    \begin{tabular}{p{8em}p{35em}c}
    \hline
    \textbf{Type} & \textbf{Tools} & \textbf{Counts} \\
    \hline
    \rowcolor{lightgrey} \multicolumn{3}{l}{Requirements Processing} \\
    \hline
    Open Source & SequenceDiagram2ESG \cite{ekici2023automatic} (Java) , spaCy \cite{gropler2021nlp} (Python), ASSERT™ tool chain \cite{moitra2019automating} (Not specified), Text2Test \cite{bhatt2022requirements} (Not specified), USLTG \cite{hue2019usltg} (Java), jUCMNav \cite{boulet2015towards, kesserwan2023transforming} (Java), SystemCockpit \cite{aichernig2014integration} (Not specified), MoMuT::REQs \cite{aichernig2014integration} (Not specified), ROO \cite{intana2023approach} (Java), ArgoUML \cite{sun2009tsgen} (Not specified), AToM3 \cite{hettab2015graph, hettab2013automatic} (Python), Sahi \cite{boghdady2011proposed} (Not specified), Semiformalizer \cite{aoyama2020test} (Not specified)  & 13  \\
    \hline
    Research Prototype  & GUI-Test tool \cite{granda2021towards} (Java), DODT \cite{aichernig2014integration} (Not specified), RETNA \cite{boddu2004retna} (Java), XML-based tool for State/Event Tree \cite{tsai2003scenario} (Java), TestMEReq \cite{moketar2016testmereq} (Not specified), CNLParser \cite{carvalho2014model} (Not specified), IMR-Generato \cite{carvalho2014model} (Not specified), CAT \cite{lu1994test} (Not specified), VDM \cite{dick1993automating} (Not specified) & 9 \\
    \hline
    Commercial  & IBM Engineering Requirements Management DOORS \cite{moitra2019automating} (Java), IBM Rational Rose \cite{kaur2018automatic, kansomkeat2003automated} (Java), Magic Draw UML \cite{singh2015functional, sarma2007automatic2} (Java)  & 3 \\
    \hline
    \rowcolor{lightgrey}\multicolumn{3}{l}{Test Generation} \\
    \hline
    Open Source & UMTG toolset \cite{wang2020automatic2} (Java), Test Generator for Cause Effect Graphs (TOUCH) \cite{krupalija2022forward} (Java), TESTSD2EFSM \cite{rocha2021model} (Java), AutomTest \cite{fernandes2020towards} (Java), TCGen \cite{perez2018automatic} (Java), SPECMATE \cite{fischbach2020specmate} (Java), Test-o-Matic \cite{masud2017automated} (Java), REBATE (REquirements Based Automatic Testing Engine) \cite{sinha2016automatic} (Java), AndroMDA \cite{elallaoui2016automatic} (Java), FDR \cite{nogueira2019test} (C++), TaRGeT \cite{nogueira2019test} (Java), Use Case Acceptance Tester (UCAT) \cite{el2010developing} (Java), SMT solver Z3 \cite{aichernig2015scalable} (C++), EclEmma \cite{malik2009scenario} (Java), T-VEC tool \cite{carvalho2014nat2testscr} (Not specified), TestGen \cite{gutierrez2008case} (Java), JWebUnit \cite{gutierrez2008case} (Java) & 17  \\
    \hline
    Research Prototype  & MC/DC-TG-RT \cite{aoyama2020test} (Java), ReqOntoTesGen \cite{intana2023approach} (Java), Tsgenerator \cite{yang2021generating} (Java), SYNTest \cite{intana2020syntest} (JavaScript), REQCAP \cite{daniel2018natural} (Not specified), SAGA \cite{daniel2018natural} (Not specified), Virtual Test Engineer (VTE) \cite{anbunathan2016automatic} (Java), ConcurTester \cite{sun2016transformation} (Java), ACT (Abstract to Concrete Tests) \cite{bubna2016act} (Java) , MISTA \cite{mirza2018automated} (Java), UBTCG \cite{oluwagbemi2015automatic} (Java), NAT2TEST \cite{carvalho2015nat2test} (Java), BT Analyser \cite{lindsay2015automation} (Not specified), C\&L (Scenarios \& Lexicons) \cite{sarmiento2014c} (Lua language), Litmus \cite{dwarakanath2012litmus} (.NET), SOLIMVA \cite{de2017statecharts} (Not specified), STATEST \cite{chatterjee2010prolific} (Not specified), ATCGT \cite{chen2010automated} (Not specified), Comprehensive Test (ComTest) \cite{swain2010test2} (Java), UTCG \cite{swain2010test} (Java), TSGen \cite{sun2009tsgen} (Java), ASSIST \cite{sarma2009automatic} (Java), GTSC \cite{santiago2012generating} (Not specified), TAD \cite{noikajana2008web} (C\# ), TDE/UML \cite{hasling2008model} (Not specified), UML behavioural test case generator (UTG) \cite{samuel2009slicing, samuel2008automatic, samuel2008novel} (Java), GenTCase \cite{ibrahim2007automatic} (C++), Condado \cite{santiago2006practical} (Not specified), UC-SCSystem \cite{nebut2006automatic} (Java), TSGAD \cite{xu2005using} (Not specified), LEIRIOS Test Generator (LTG) \cite{bouquet2005requirements} (Not specified), ScenTED-DTCD tool \cite{reuys2005model} (Not specified), UMLTGF \cite{linzhang2004generating} (Not specified), AGATHA \cite{lugato2004automated} (Not specified), EXTRACT \cite{yu2003generating} (Not specified), UCTSystem \cite{nebut2003requirements} (Java), UMLTest \cite{offutt1999generating} (Not specified), Objecteering CASE \cite{nebut2003requirement} (Not specified), PerformCharts \cite{santiago2006practical} (Not specified), UsageTester \cite{riebisch2002uml} (Not specified), TCaseUML \cite{sun2008transformation} (Not specified), Requirements Model Code Synthesis (RMCS) \cite{gupta2001synthesis} (Not specified), RT-Tester \cite{carvalho2014model} (Not specified) & 43  \\
    \hline
    Commercial  & Spec explorer \cite{yazdani2019automatic} (C\#), Conformiq Creator \cite{hussain2018modeling} (Java), Conformiq’s Qtronic \cite{abbors2009tracing} (Java)  & 3 \\
    \hline
    \end{tabular}
\end{table}

\begin{figure}[!t]
  \centering
  \subfigure[Tools types.]{
  \label{fig:Tools1}
  \includegraphics[width=0.46\linewidth]{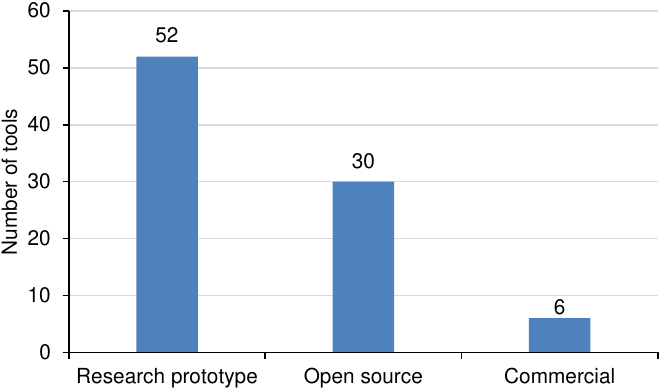}}
  \subfigure[Tools' development language.]{
  \label{fig:Tools2}
  \includegraphics[width=0.49\linewidth]{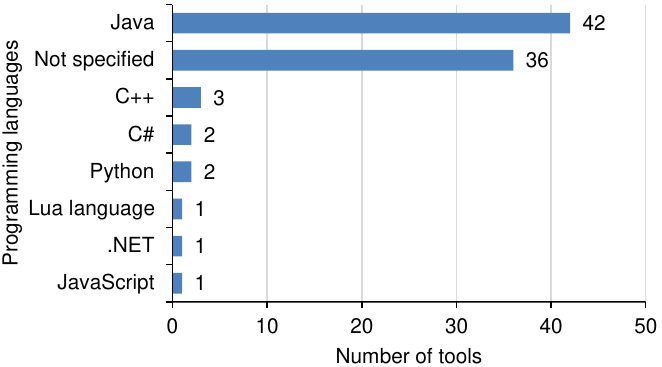}}
  \caption{Distribution of RBTG tools}
  \Description{Distribution of RBTG Tools.}
  \label{fig:tools}
\end{figure}

In the requirements-processing category, open source tools are ranked highly (13 tools), followed by research-prototype tools (nine ) and commercial tools (three). 
In the test-generation category, research-prototype tools lead significantly (43), with open-source (17) and commercial (three) tools being less common.

Overall, as shown in Figure \ref{fig:Tools1}, research-prototype tools were the most common RBTG tools, followed by the open source, and then commercial tools.  
Regarding the programming languages used in the implementation of these tools, as shown in Figure \ref{fig:Tools2}, Java stands out as the most widely used, accounting for 42 tools. 
36 tools did not have their implementation languge specified.
The other languages (C++, C\#, Python, Lua, .NET, and JavaScript), were used in significantly fewer cases. 

\section{Answer to RQ6: How Have RBTG Approaches Been Evaluated?
\label{section:RQ6}}

\subsection{Overall Evaluation in RBTG}

\citet{chen2011systematic} described nine types of evaluation methods for software product lines. 
Four of these were analyzed statistically based on the papers reviewed in this study. 
To ensure clarity, we renamed the evaluation types as \textit{Example}, \textit{Empirical Case Study}, \textit{Real-World Case Study}, and \textit{Discussion}.
\begin{itemize}
  \item {\texttt{Example}}.
  In the example category, an example is used to demonstrate how the proposed method works. The example is primarily illustrative and not intended to validate or evaluate the approach. For instance, 
  \citet{ueda2024automatic} provided an example covering the entire process from requirements specification through data processing to test-case generation and evaluation. 
  \citet{yonathan2024generating} used the login feature as an example to illustrate both the process and the results. 
  \citet{hong2001automatic} did the same using a coffee-vending machine scenario.
  
  \item {\texttt{Empirical Case Study}}. 
  This explores a phenomenon within its real-life context. 
  While based on real scenarios, these studies are typically conducted in controlled academic settings, and explicitly designed as case studies.
  \citet{gupta2023decision} demonstrated their technique using a case study of ATM cash-withdrawal processes.
  
  \item {\texttt{Real-World Case Study}}. 
  In this category, approaches are applied to real-world systems, 
  often in collaboration with companies or public organizations. 
  These studies emphasize practical applicability, and evidence of success in operational environments.
  \citet{arora2024generating} collaborated with Austrian Post, an international logistics and service provider, 
  to generate test scenarios from NL requirements. 
  \citet{intana2023approach} demonstrated the effectiveness and accuracy of their approach and tool through two case studies involving real-world systems.
  
  \item {\texttt{Discussion}}. 
  Discussions provide some qualitative, textual, and opinion-based evaluation, 
  without concrete examples or empirical data.
  \citet{elqortobi2023granular} and \citet{lafi2021automated}, for example, analyzed the test-generation process primarily through discussions, without providing illustrative examples.
\end{itemize}

\begin{figure}[htbp]
  \centering
  \subfigure[Empirical evaluations.]{
  \label{fig:evaluation1}
  \includegraphics[width=0.48\linewidth]{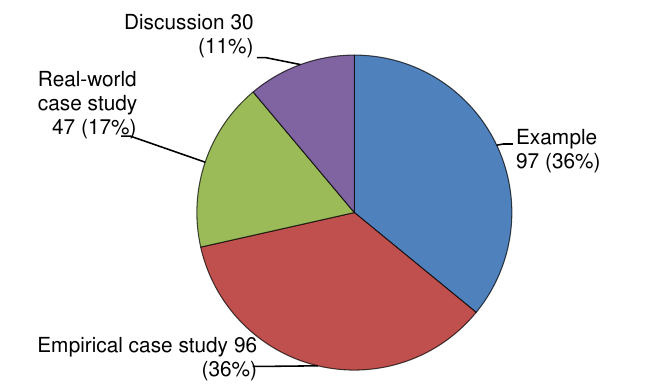}}
  \subfigure[Evaluation methods.]{
  \label{fig:evaluation2}
  \includegraphics[width=0.48\linewidth]{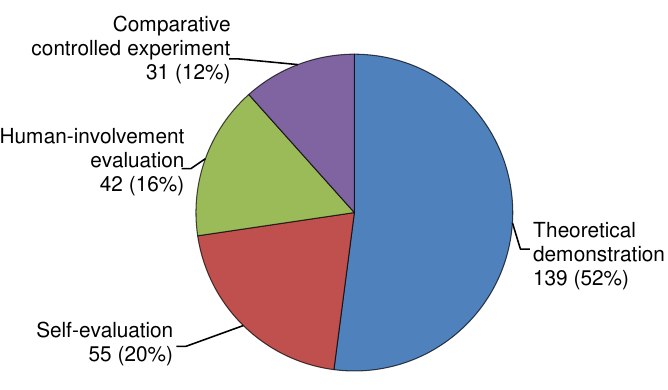}}
  \caption{Overall distribution of evaluation methods.}
  \Description{.}
  \label{fig:evaluation}
\end{figure}

The 267 studies were classified based on their primary evaluation methods. 
Figure \ref{fig:evaluation1} shows that examples and empirical case studies were the largest proportion, each accounting for 36\% of the total. 
In contrast, only 17\% of the studies used real-world case studies \cite{schnelte2009generating}, and 11\% were categorized as discussions.  

We categorized the reviewed studies according to their validation methods into four distinct types: 
\textit{theoretical demonstration}, \textit{self-evaluation}, \textit{human-involvement evaluation}, and \textit{comparative controlled experiment}.
\begin{itemize}
  \item {\texttt{Theoretical Demonstration}}. 
  This category included studies that primarily introduce and illustrate  methods through theoretical explanations and examples, without any formal evaluation. 
  \citet{chen2023research}, \citet{kaur2014generation}, and \citet{reuys2003derivation} presented their test-generation process without conducting any evaluation.
  
  \item {\texttt{Self-Evaluation}}. 
  This category included research that performs an initial assessment of the proposed approach using relevant data, typically focusing on internal validation without comparison to other methods. 
  \citet{gupta2023decision} and \citet{thanakorncharuwit2016generating} evaluated their generated test cases based on some types of test coverage, while \citet{wang2019specification} involved mutation testing in the evaluation.
  
  \item {\texttt{Human-Involvement Evaluation}}. 
  This category involved studies incorporating human participation, either through manual assessment of  results, or by comparing automated outputs with expert-generated ones. 
  \citet{arora2024generating}, \citet{shah2019test}, and \citet{fernandez2012practical} had their proposed approach critically evaluated by experts. \citet{ekici2023automatic} compared their automatically generated test sequences with those created manually.
  
  \item {\texttt{Comparative Controlled Experiment}}.
  The final category included research that had controlled experiments comparing the proposed method against existing approaches.
  This category provided rigorous validation of the proposed methods' effectiveness and advantages. 
  \citet{agrawal2020scaling} conducted experiments with three other algorithms, using corresponding parameter values for comparison. 
  \citet{elghondakly2016optimized} and \citet{boghdady2012automatic} 
  computed cyclomatic complexity values to measure the number of logical paths in a module and compared these values with those obtained from other methods.
\end{itemize}

Figure \ref{fig:evaluation2} shows that theoretical demonstration was the most common validation method, accounting for 52\% of the studies. 
Self-evaluation, human-involvement evaluation, and comparative controlled experiments all had similar representation, with 20\%, 16\%, and 12\% of the studies, respectively. 
Notably, this suggests that over half of the reviewed studies lack formal evaluation.

In summary, the dominance of examples and empirical case studies indicates that many proposed approaches lack sufficient practical validation:
Only 17\% of the studies employed case studies based on real-world systems, with most relying on theoretical demonstrations. 
Only 12\% of the studies conducted comparative controlled experiments. 
The limited use of formal validation methods highlights a significant gap between theoretical research and practical applicability, making it difficult for practitioners to rely solely on existing research for decision making.

\subsection{Evaluation Methods and Metrics in RBTG}

Various methods have been applied to evaluate proposed RBTG approaches.
A range of metrics have been used to measure the efficiency and effectiveness of the generated test cases. 
This section provides a comprehensive overview of these evaluation methods, including the key metrics used in the reviewed studies.

\begin{figure}[h]
  \centering
  \includegraphics[width=0.7\linewidth]{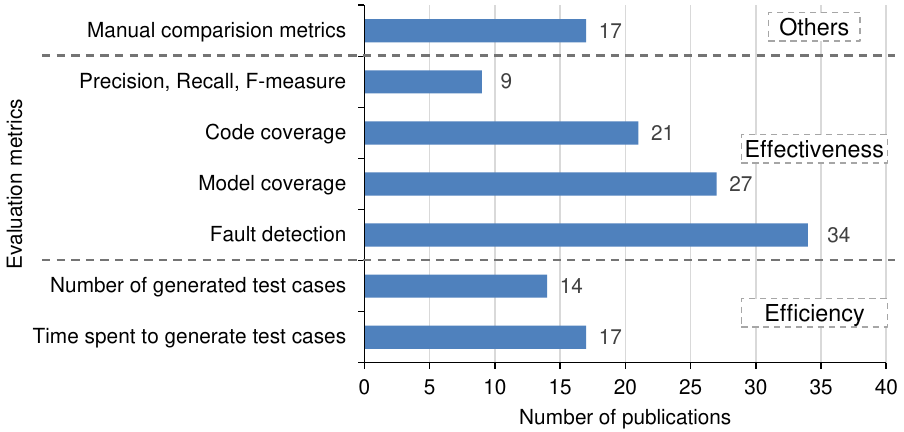}
  \caption{Distribution of evaluation metrics.}
  \Description{Evaluation metrics.}
  \label{fig:evaluation3}
\end{figure}

Figure \ref{fig:evaluation3} summarizes the metrics adopted in the surveyed literature, categorized into three groups: 
\textit{Efficiency}, \textit{Effectiveness}, and \textit{Other}. 
Similar metrics, such as fault detection, code coverage, and model coverage, were grouped together within the categories. 
A single paper may have used multiple metrics in the evaluations.

The {\em effectiveness} metrics were categorized into four types: 
\textit{fault detection}, \textit{model coverage}, \textit{code coverage}, and \textit{precision, recall, and F-measure}. 
\begin{itemize}
    \item {\texttt{Fault detection}}. 
    This metrics include \textit{mutation score} \cite{liu2020automatic, xu2014automated, kansomkeat2010generating} and \textit{detected faults} \cite{ekici2023automatic, kundu2009novel}. 
    The mutation score, calculated as the ratio of killed mutants to the total number of mutants, is a widely used measure in mutation testing \cite{liu2020automatic}. 
    The detected faults metric quantifies the number of defects identified by the generated test cases, providing a direct assessment of their fault-detection capability. 
    
    \item {\texttt{Model coverage}}. 
    This category includes metrics such as \textit{transition coverage} \cite{gupta2023decision}, \textit{cyclomatic complexity} \cite{yonathan2024generating, chouhan2012test}, \textit{activity coverage} \cite{gupta2023decision}, \textit{action coverage} \cite{ali2021model}, and \textit{fitness value} \cite{lakshminarayana2020automatic}. 
    It also includes other coverage criteria, such as \textit{state coverage} \cite{pradhan2022transition} and \textit{key path coverage} \cite{chen2010efficient}. 
    Coverage measurement, which evaluates whether or not the model can reach a certain state or transition, is a commonly used evaluation method.
    
    \item {\texttt{Code coverage}}. 
    This category includes metrics such as \textit{path coverage} \cite{gupta2023decision}, \textit{MC/DC coverage} \cite{wang2022mc, zheng2021generating, rayadurgam2001test}, \textit{branch coverage} \cite{wang2020automatic2}, and \textit{row coverage} \cite{agrawal2020scaling}.
    It also includes other coverage criteria, like \textit{predicate coverage} \cite{swain2010test2}. 
    Code coverage measurements have also been frequently used.
    
    \item {\texttt{Precision, recall, F-measure}}. 
    The \textit{F-measure}, also known as the F1 Score, combines precision and recall and is used to measure the accuracy of automatic test-case generation \cite{ueda2024automatic}. 
    The formula for the F-measure is as follows: 
        \begin{equation}
            {Precision} = \frac{{TP}}{{TP} + {FP}}
        \end{equation}
        
        \begin{equation}
            {Recall} = \frac{{TP}}{{TP} + {FN}}
        \end{equation}
        
        \begin{equation}
            {F-measure} = \frac{2 \times {Precision} \times {Recall}}{{Precision} + {Recall}}
        \end{equation}
    (where TP is the True Positive, FP is the False Positive, and FN is the False Negative). 
\end{itemize}

The evaluation of RBTG testing {\em efficiency} primarily used two key metrics: 
\textit{time spent to generate test cases} \cite{ali2021model}, and \textit{number of generated test cases} \cite{chen2022improved}.
These metrics are commonly used to assess the reduction in manual effort needed for effective test-case generation.

Manual comparison metrics \cite{arora2024generating} involve human evaluation based on criteria such as relevance, coverage, correctness, understandability, and feasibility. 
This can provide qualitative insights into the effectiveness of the generated test cases, complementing quantitative evaluation approaches.

Other studies also adopted specialized evaluation methods, such as comparisons with manual evaluation (cross-validated by two evaluators) \cite{lim2024test} and proof-of-concept evaluation \cite{daniel2018natural}, to assess the feasibility and effectiveness of the proposed approaches.

\section{Answer to RQ7: In What Domains and Industrial Applications Has RBTG Been Applied?
\label{section:RQ7}}

The SUT application domain strongly influences RBTG, impacting the formulation and representation of requirements, and the selection of test-generation approaches and tools. 
Test cases can differ significantly between domains in terms of number, style, detail, and granularity. 
This section identifies the key application domains that have been explored for RBTG.

\begin{figure}[!b]
  \centering
  \includegraphics[width=0.8\linewidth]{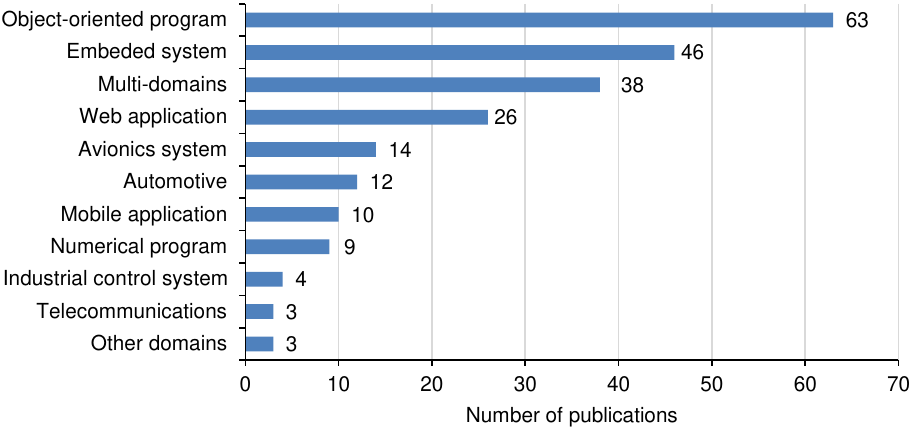}
  \caption{Distribution of RBTG application domains.}  
  \Description{Distribution of domains in RBTG.}
  \label{fig:domain}
\end{figure}

228 of the 267 reviewed papers explicitly referenced specific application domains:
These were incorporated into the analysis. 
For papers lacking explicit domain specifications, their case studies' domains were used.
The resulting distribution of application domains is presented in Figure \ref{fig:domain}. 
The most extensively studied domain was object-oriented programs, in 63 papers.  
Other notable domains were embedded systems (46 papers), multi-domains (38 papers), and web applications (26 papers). 
The term multi-domain refers to approaches applied across multiple system types, such as both object-oriented programs and embedded systems \cite{boghdady2012enhanced, boghdady2011enhanced}.
This highlights the significance of object-oriented programs in RBTG research, 
with an emphasis on model-based approaches that typically employ UML diagrams to describe object-oriented systems.

In object-oriented programs, objects interact to execute specific behaviors \cite{kaur2018automatic}. 
Within the context of RBTG research, a broad range of applications have been modeled as object-oriented programs, including for:
library management \cite{intana2023approach, intana2020syntest, briand2002uml}, online conference reviewing \cite{granda2014towards}, kidney-failure diagnosis (KFD) \cite{intana2023approach}, university education  \cite{yazdani2019automatic}, and course registration \cite{kumaran2011approach}. 
This diversity in domains highlights the flexibility and adaptability of object-oriented modeling for RBTG.

Embedded systems, which typically operate with continuous digital input and output signals, have also been extensively explored in RBTG surveys. 
In such systems, input data are generally collected from sensors as the primary data source, and output data are transmitted to system actuators to perform control actions \cite{silva2019cpn}. 
Examples of embedded systems studied in RBTG research include systems for elevator control \cite{chen2023research}, urban railway interlocking \cite{kadakolmath2022model},
BodySense (a safety-critical automotive software) \cite{wang2020automatic2}, railroad crossings \cite{krupalija2022forward} and medical equipment \cite{tahat2001requirement}. 
These examples highlight the key role of embedded systems across safety-critical and control-oriented applications.

Web applications have also been a focus of RBTG research due to their widespread use in real-world systems. 
Example web applications include:
the automatic exam-paper evaluation platform AutoEval \cite{mathur2023automated};
the online trading platform Online Broker \cite{sarmiento2016test}; and 
the hotel-management system Hotel Ambassador \cite{gutierrez2015model}. 
These studies highlight the applicability of RBTG to software systems with the kind of dynamic and user-driven functionality typical of web-based applications.

Avionics and automotive domains have also drawn considerable attention in RBTG research, with 14 and 12 papers, respectively. 
Some examples in these domains include the Avionics Launch Abort System \cite{wang2022mc}, Aircraft Engine Sensor \cite{moitra2019automating} and Adaptive Cruise Control System \cite{aniculaesei2018automated}.

Several other application domains have also been explored:
Mobile applications, such as the Uber app \cite{ali2021model}, are linked to 10 studies. 
Numerical programs, including the \textit{greatest common divisor} (GCD) \cite{cajica2021automatic} and Mod process \cite{wang2019specification}, were examined in nine papers. 
Industrial control systems \cite{crapo2019using} and telecommunications \cite{kikuma2018automatic} have been explored in four and three papers, respectively. 
Other less explored domains, such as healthcare \cite{loffler2010formal}, desktop applications \cite{granda2021towards}, and logistics  \cite{arora2024generating} 
were each covered in a single paper. 
These numbers highlight the applicability of RBTG across a wide range of industries and use cases.

Overall, while object-oriented programs, embedded systems, and web applications dominate the RBTG research, the diversity of applications highlights the adaptability of RBTG approaches across different domains. 
However, the limited exploration of some domains suggests opportunities for further research to enhance test-case generation techniques in these underrepresented areas.

\section{Answer to RQ8: What Are the Remaining Challenges and Future Work for RBTG?
\label{section:RQ8}}

Following the analyses in the previous sections, this section summarizes remaining challenges and potential future work for RBTG.

\subsection{Challenges}

This survey paper has identified that RBTG has attracted considerable research interest and has made notable progress in recent years.
However, compared to some fields (like automated code generation), its advances remains limited, and the level of automation achieved is still lacking. 
Many challenges and open questions remain to be addressed.

\subsubsection{Challenge 1: Constructing High-Quality Requirements}

Requirements serve as the foundation of RBTG, directly influencing the strategies for test-case generation, and determining the quality of the resulting test cases. 
In industrial practice, most requirements are expressed in NL, which can face lexical ambiguities, contextual inconsistencies, and structural incompleteness \cite{mustafa2021automated}:
Test cases thus generated are often unreliable, significantly impacting the overall quality of the software testing and development. 
In traditional development processes, ensuring the precision, completeness, clarity, and consistency of the requirements often involves iterative reviews, conducted manually by engineers during the early project stages. 
The requirements can then be continuously refined throughout the development and testing phases to maintain their quality and adaptability.  

The reviewed RBTG research revealed that efforts have been made to mitigate NL-related issues by translating into semi-formal \cite{aoyama2020test} or formal representations, or through mapping tables for NLP \cite{lafi2021automated}. 
However, these approaches often require substantial manual effort \cite{granda2021towards}, risk losing important information (such as non-functional requirements), and do not guarantee that the requirements will be complete and unambiguous.

To address these limitations, a potential research direction involves integrating requirements engineering with automated testing to establish high-quality, machine-readable requirements for test-case generation. 
This could involve the use of prototypes, scenarios, stories, and domain knowledge to construct comprehensive and precise requirements. 
Advances in NLP and LLMs may also offer potential solutions for refining NL requirements.

\subsubsection{Challenge 2: From Abstract to Executable Test Cases}

CTCs are executable, and their automatic generation represents the ultimate goal of RBTG.
However, the surveyed studies reveal that RBTG primarily produces ATCs. 
These generated test cases are often incomplete, and lack sufficient detail:
This makes them unready for direct execution. 
To address this, researchers have explored methods to convert ATCs into CTCs. 

\citet{wang2020automatic2} used a mapping table to develop an executable test generator that can automatically produce CTCs. 
The table maps high-level operation descriptions and test inputs to concrete driver functions and inputs. 
\citet{maciel2019requirements} refined ATCs by assigning values to entities and creating temporary variables.
These were then transformed into concrete test scripts by establishing relations between the requirements and the syntax of the Robot framework \cite{gupta2011model}, using keywords from the Selenium library. 
\citet{kesserwan2019use} semi-automatically mapped abstract \textit{Test Description Language} (TDL) datasets to concrete data, generating executable test procedures in TTCN-3 \cite{kesserwan2023transforming} for embedded systems. 

A common approach for the creation of concrete test data and oracles involves random generation or manual crafting. 
\citet{fernandes2020towards} proposed AutomTest, a tool that enables users to generate automated tests from functional requirements:
It accepts data within parameter ranges and produces JUnit test cases. 
However, its applicability is constrained by limited variable support and the need for manual intervention. 
\citet{pradhan2022transition} developed test cases using TestNG:
automatically, based on their specific conditions.
\citet{moitra2019automating} developed executable test scripts tailored for safety-critical systems:
These tests can be run using \textit{Safety-critical application development environment} (SCADE) coverage tools. 
However, these approaches are often based on formal requirements, and constrained to specific domains.

Exploring a general approach that could automatically convert ATCs to CTCs remains a significant challenge:
Addressing these issues will be essential for advancing RBTG.

\subsubsection{Challenge 3: Lack of Standard Benchmarks}

Research benchmarks enable robust analysis and fair comparisons of automated test-generation approaches. 
They typically include systems specifically designed for a particular type of evaluation, facilitating comparisons with other research, reducing system-selection bias, and demonstrating the effectiveness of proposed techniques \cite{fontes2023integration}. 
Currently, the ATM system \cite{sahoo2021test} is the most widely-used benchmark. 
However, the majority of surveyed studies rely only on their own case studies, with nearly half of the reviewed research not involving any case-study validation.  

The absence of standardized benchmarks makes it difficult to compare different approaches, reproduce results, and assess real-world applicability. 
To address this gap, well-documented benchmarking datasets should be developed, incorporating diverse requirements formats (e.g., NL, semi-formal, and model-based).  
These benchmarks should cover various domains (including embedded systems and enterprise applications) to help ensure consistent evaluation, and support progress in RBTG.  

\subsubsection{Challenge 4: Rigorous and Efficient Evaluation}

Figure \ref{fig:evaluation2} showed that only 12\% of the reviewed studies included controlled comparative experiments, and 16\% required human involvement, introducing subjectivity into the evaluation. 
Current evaluation metrics (such as code coverage, model coverage, and fault-detection rates) may only provide a limited perspective on the effectiveness of RBTG approaches: 
They may fail to capture how well generated test cases align with the original requirements, especially for non-functional requirements such as performance, security, and usability.

The lack of a comprehensive evaluation framework hinders efforts to determine whether or not generated test cases adequately capture the intended SUT behaviors.
To advance RBTG research, rigorous evaluation methodologies are required. 
These methodologies should extend beyond traditional metrics, incorporating both functional and non-functional dimensions, to enable a more comprehensive assessment of test-case generation techniques.

\subsubsection{Challenge 5: Real-World Industrial Implementation}

Despite advances in RBTG, its adoption in industrial settings remains limited. 
As discussed in Section \ref{section:RQ6}, only 17\% of the reviewed papers used real-world case studies to evaluate RBTG, and these often spanned different domains. 
\citet{arora2024generating} applied their approach to two projects at Austrian Post, where domain experts confirmed the coherence, feasibility, and comprehensiveness of the generated test scenarios. 
However, the necessity for human-expert involvement to confirm correctness was also noted. 

This challenge highlights the barriers to wider adoption and scaling-up of RBTG methods in industrial settings. 
Addressing these will require that RBTG approaches align with practical needs, are cost-effective, and can adapt to domain-specific constraints.

\subsection{Future work}

The challenges identified above should be addressed through more effective and efficient approaches, which should be considered as future RBTG work. 
This section lists some potential future work for RBTG.

\subsubsection{Future Work 1: Exploring LLM-Driven RBTG}

A promising future direction will be the application of LLMs to RBTG. 
Current methods often struggle with ambiguous or incomplete requirements, leading to suboptimal test-case generation. 
Using the advanced NLP capabilities of LLMs may enable more accurate parsing and clarification of these ambiguous requirements. 
Once clarified, formal methods could then be applied to generate precise constraints.
This would have the potential to significantly enhance the quality and effectiveness of the generated test cases. 
Further research in this area should also explore how LLMs could be integrated into existing RBTG frameworks, evaluating their performance in real-world industrial settings.

\subsubsection{Future Work 2: Optimizing Module-Centric Test Prioritization}

Optimizing module-centric test prioritization is another promising future direction. 
Identifying characteristics of error-prone modules during requirements analysis and test-case design, especially within specific environments, can provide valuable insights and improve testing effectiveness.
Since 80\% of defects reportedly come from 20\% of modules \cite{boehm2007software}, the critical modules should be tested more thoroughly.

Although previous research has explored the optimization of test cases through removal of redundant ones \cite{intana2023approach}, few studies have adjusted the number of generated test cases 
based on module criticality.
However, it may also sometimes be necessary to tolerate a certain level of redundancy when testing critical modules, to ensure comprehensive testing.

\subsubsection{Future Work 3: Engineering Effective RBTG Tools}

Despite the abundance of research on RBTG tools, their adoption in industry remains limited. 
This may be due to a lack of clear evidence that these tools provide significant cost savings or benefits over traditional methods. 
Developing effective and user-friendly test-generation tools, therefore, is another crucial future direction. 
This will help bridge the gap between academic research and practical applications, providing industry with useful tools.

\subsubsection{Future Work 4: Exploring Non-Functional RBTG}

In the reviewed studies, RBTG was predominantly used for functional testing, with limited applications in performance \cite{ali2021model}, security \cite{chen2022improved}, and other non-functional requirements. 
The primary reason for this is that functional requirements constitute the majority of documented requirements. 
Additionally, performance testing and security testing often require more specialized models that detect and analyze specific types of data \cite{wang2024software}.

Despite the availability of performance tools like LoadRunner \cite{loadrunner} and JMeter \cite{jmeter}, there remains a need for manual script design. 
Future work should integrate these tools to automate non-functional testing, reducing manual effort and improving efficiency.

\section{Conclusion
\label{section:Conclusion}}

This paper has provided a comprehensive survey of 267 \textit{requirements-based test generation} (RBTG) studies published up to 2024. 
We traced the distribution of RBTG topics, and analyzed the entire process of generating test cases based on requirements. 
We classified input requirements' types and approaches; 
identified different types of test cases; and 
conducted a statistical review of existing tools and evaluation methods. 
We have also highlighted the diverse applications of RBTG across multiple domains, and discussed the challenges and potential opportunities in the field.

Our review revealed that early RBTG methods primarily used formal or semi-formal requirements' representations to generate test cases, with model-based approaches emerging as the primary strategy. 
However, recent advancements in \textit{natural language processing} (NLP) have shifted the focus towards NL requirements. 
A significant amount of previous research focused on model-based testing, which faces challenges for complex systems. 
There was also a notable lack of methods for transforming \textit{abstract test cases} (ATCs) into \textit{concrete test cases} (CTCs), and a lack of effective evaluation methods.

There are several RBTG challenges that require future investigation. 
First, input requirements are often not fully integrated with the domain knowledge, the prototype, and other information. 
Secondly, most test cases generated by RBTG are ATCs, with few universal methods for converting ATCs into CTCs. 
Finally, current evaluation processes remains insufficient. 
There is also a lack of general-purpose automated testing tools 
that can substantially reduce manual effort.
We have also identified some potential opportunities, such as the application of \textit{large language models} (LLMs) to RBTG, 
potentially enhancing test generation for critical modules.

This paper provides a comprehensive overview of the current state, gaps, and future research directions for RBTG.
With automated code generation already widely adopted in industry, the demand for automated testing is more urgent than ever.
The insights provided in this paper will benefit both researchers and practitioners, improving testing methodologies and ultimately enhancing the quality and reliability of software systems.

\section*{Acknowledgment}

This work in this paper was partly supported by the Science and Technology Development Fund of Macau, Macau SAR, under Grant No. 0021/2023/RIA1.

\bibliographystyle{ACM-Reference-Format}
\bibliography{survey_ref}

\appendix

\section{List of Surveyed Studies}

Table \ref{tab:listofsurveyedstudies} presents the finalized list of 267 research studies surveyed in this work.

\renewcommand{\thetable}{A.\Roman{table}}
\setcounter{table}{0}
\begin{scriptsize}
\begin{longtable}{m{0.5cm}|m{4.5cm}|m{2cm}|m{4.5cm}|c}
\caption{List of Surveyed Studies.}
\label{tab:listofsurveyedstudies} \\
\hline
\textbf{No.} & \textbf{Title} & \textbf{Authors} & \textbf{Venue} & \textbf{Year} \\ \hline
\endfirsthead

\caption{(Continued).}\\
\hline
\textbf{No.} & \textbf{Title} & \textbf{Authors} & \textbf{Venue} & \textbf{Year} \\ \hline
\endhead
1 & Test case generation for   specification-based software testing & Peng & Proceedings of the 1994   Conference of the Centre for Advanced Studies on Collaborative Research & 1994 \\ \hline
2 & A framework for   specification-based testing & Stocks and Carrington & IEEE Transactions on Software   Engineering & 1996 \\ \hline
3 & Testing from a Z Specification & Hierons RM & Software Testing, Verification   and Reliability & 1997 \\ \hline
4 & Specification‐based testing   using cause‐effect graphs & Paradkar et al. & Annals of Software Engineering & 1997 \\ \hline
5 & Testing object-oriented   programs: from formal specification to test scenario generation & Chang et al. & Journal of Systems and Software & 1998 \\ \hline
6 & Test scenario generation from a   structured requirements specification & Cunning and Rozenbiit & Proceedings ECBS'99. IEEE   Conference and Workshop on Engineering of Computer-Based Systems & 1999 \\ \hline
7 & Using model checking to generate   tests from requirements specifications & Gargantini and Heitmeyer & ACM SIGSOFT Software Engineering   Notes & 1999 \\ \hline
8 & A scenario-based approach to   validating and testing software systems using statecharts & Ryser and Glinz & Proceedings of the 12th   International Conference on Software and Systems Engineering and their   Applications & 1999 \\ \hline
9 & Automated test case generation   from dynamic models & Fr{\"o}hlich and Link & Proceedings of the European   Conference on Object-Oriented Programming & 2000 \\ \hline
10 & UML-Based   integration testing & Hartmann et al. & Proceedings of the 2000 ACM   SIGSOFT international symposium on Software testing and analysis & 2000 \\ \hline
11 & Automatic test generation from   statecharts using model checking & Hong et al. & Workshop on Formal Approaches to   Testing of Software & 2001 \\ \hline
12 & Synthesis of high-level   requirements models for automatic test generation & Gupta et al. & Proceedings of the 8th Annual   IEEE International Conference and Workshop on the Engineering of   Computer-Based Systems & 2001 \\ \hline
13 & Test-sequence generation from   formal requirement models & Rayadurgam and Heimdahl & Proceedings of the 6th IEEE   International Symposium on High Assurance Systems Engineering & 2001 \\ \hline
14 & Automated generation of   statistical test cases from UML state diagrams & Chevalley and Thevenod-Fosse & Proceedings of the 25th Annual   International Computer Software and Applications Conference & 2001 \\ \hline
15 & Requirement-based automated   black-box test generation & Tahat et al. & Proceedings of the 25th annual   international computer software and applications conference & 2001 \\ \hline
16 & Generation of functional test   sequences from B formal specifications presentation and industrial case-study & Legeard and Peureux & Proceedings 16th Annual   International Conference on Automated Software Engineering & 2001 \\ \hline
17 & Dependence analysis in reduction   of requirement based test suites & Vaysburg et al. & Proceedings of the 2002 ACM   SIGSOFT International Symposium on Software Testing and Analysis & 2002 \\ \hline
18 & A UML-based approach to system   testing & Briand and Labiche & Software and Systems Modeling & 2002 \\ \hline
19 & Derivation of domain test   scenarios from activity diagrams & Reuys et al. & Proceedings of the International   Workshop on Product Line Engineering & 2003 \\ \hline
20 & A requirement-based approach to   test product families & Nebut et al. & Proceedings of the 2003   International Workshop on Software Product-Family Engineering & 2003 \\ \hline
21 & UML-based statistical test case   generation & Riebisch et al. & Proceedings of the 2002   International Conference on Objects, Components, Architectures, Services, and   Applications for a Networked World & 2003 \\ \hline
22 & Generating tests from UML   specifications & Offutt and Abdurazik & Proceedings of the 1999   International Conference on the Unified Modeling Language & 2003 \\ \hline
23 & Scenario-based test case   generation for state-based embedded systems & Tsai et al. & Proceedings of the 2003 IEEE   International Performance, Computing, and Communications Conference & 2003 \\ \hline
24 & Automated-generating test case   using UML statechart diagrams & Kansomkeat and Rivepiboon & Proceedings of the 2003 annual   research conference of the South African institute of computer scientists and   information technologists on Enablement through technology & 2003 \\ \hline
25 & Requirements by contracts allow   automated system testing & Nebut et al. & Proceedings of the 14th   International Symposium on Software Reliability Engineering & 2003 \\ \hline
26 & Generating, selecting   and prioritizing test cases from specifications with tool support & Yu et al. & Proceedings of the 3rd   International Conference on Quality Software & 2003 \\ \hline
27 & Formal test-case generation for   UML statecharts & Gnesi et al. & Proceedings of the 9th IEEE   International Conference on Engineering of Complex Computer Systems & 2004 \\ \hline
28 & Automated functional test case   synthesis from THALES industrial requirements & Lugato et al. & Proceedings of the 10th IEEE   Real-Time and Embedded Technology and Applications Symposium & 2004 \\ \hline
29 & RETNA: from   requirements to testing in a natural way & Boddu et al. & Proceedings of the 12th IEEE   International Requirements Engineering Conference & 2004 \\ \hline
30 & Generating test cases from UML activity diagram based on Gray-box method & Linzhang et al. & Proceedings of the 11th   Asia-Pacific Software Engineering Conference & 2004 \\ \hline
31 & Model-based system testing of   software product families & Reuys et al. & Advanced Information Systems   Engineering & 2005 \\ \hline
32 & UCM-driven testing of web   applications & Amyot et al. & Proceedings of the 12th   International SDL Forum on SDL 2005: Model-Driven Engineering for Smart   Cities & 2005 \\ \hline
33 & Automating test case generation   from Z specifications with Isabelle & Helke et al. & Proceedings of 10th the   International Conference of Z Users & 2005 \\ \hline
34 & Automating the generation and   sequencing of test cases from model-based specifications & Dick and Faivre & International Symposium of   Formal Methods Europe & 2005 \\ \hline
35 & Requirements traceability in   automated test generation: application to smart card software validation & Bouquet et al. & ACM SIGSOFT Software Engineering   Notes & 2005 \\ \hline
36 & Using adaptive agents to   automatically generate test scenarios from the UML   activity diagrams & Xu et al. & Proceedings of the 12th   Asia-Pacific Software Engineering Conference & 2005 \\ \hline
37 & Synthesis of scenario based test   cases from  B   models & Satpathy et al. & Proceedings of the 2006   International Workshop on Formal Approaches to Software Testing & 2006 \\ \hline
38 & Generating test sequences from   UML sequence diagrams and state diagrams & Sokenou & INFORMATIK 2006--Informatik   f{\"u}r Menschen--Band 2, Beitr{\"a}ge der 36. Jahrestagung der   Gesellschaft f{\"u}r Informatik eV & 2006 \\ \hline
39 & Automatic test generation: a use   case driven approach & Nebut et al. & IEEE Transactions on Software   Engineering & 2006 \\ \hline
40 & Coverage metrics for   requirements-based testing & Whalen et al. & Proceedings of the 2006   international symposium on Software testing and analysis & 2006 \\ \hline
41 & A practical approach for   automated test case generation using statecharts & Santiago et al. & Proceedings of the 30th Annual   International Computer Software and Applications Conference & 2006 \\ \hline
42 & Generating interaction test   cases for mobile phone systems from use case specifications & de Figueiredo et al. & ACM SIGSOFT Software Engineering   Notes & 2006 \\ \hline
43 & Automatic test case generation   from UML communication diagrams & Samuel et al. & Information and Software   Technology & 2007 \\ \hline
44 & Test cases generation from UML   activity diagrams & Kim et al. & Proceedings of the 8th ACIS   international conference on software engineering, artificial intelligence,   networking, and parallel/distributed computing & 2007 \\ \hline
45 & Test case generation by means of   UML sequence diagrams and labeled transition systems & Cartaxo et al. & Proceedings of the 2007 IEEE   International Conference on Systems, Man and Cybernetics & 2007 \\ \hline
46 & An automatic tool for generating   test cases from the system's requirements & Ibrahim et al. & Proceedings of the 7th IEEE   International Conference on Computer and Information Technology & 2007 \\ \hline
47 & Automatic test case generation   from UML sequence diagram & Sarma et al. & Proceedings of the 15th   International Conference on Advanced Computing and Communications & 2007 \\ \hline
48 & Automatic test case generation   from UML models & Sarma and Mall & Proceedings of the 10th   International Conference on Information Technology & 2007 \\ \hline
49 & Requirement-based testing   through formal methods & Cabral and Tamai & Proceedings of TESTCOM-FATES & 2008 \\ \hline
50 & A novel test case design   technique using dynamic slicing of UML sequence diagrams & Samuel and Mall & e-Informatica Software   Engineering Journal & 2008 \\ \hline
51 & Requirements-based test case   specification by using information from model construction & Pfaller & Proceedings of the 3rd   International Workshop on Automation of Software Test & 2008 \\ \hline
52 & An approach for supporting   system-level test scenarios generation from textual use cases & Some and Cheng & Proceedings of the 2008 ACM   symposium on Applied computing & 2008 \\ \hline
53 & Automatic test case generation   using unified modeling language (UML) state diagrams & Samuel et al. & The Institution of Engineering   and Technology & 2008 \\ \hline
54 & Model based testing of system   requirements using UML use case models & Hasling et al. & Proceedings of the 1st   International Conference on Software Testing, Verification, and Validation & 2008 \\ \hline
55 & A case study for generating test   cases from use cases & Gutierrez et al. & Proceedings of the 2nd   International Conference on Research Challenges in Information Science & 2008 \\ \hline
56 & A transformation-based approach   to generating scenario-oriented test cases from UML activity diagrams for   concurrent applications & Sun & Proceedings of the 32nd Annual   IEEE International Computer Software and Applications Conference & 2008 \\ \hline
57 & Test Sequence Generation from   UML Sequence Diagrams & Samuel and Joseph & Proceedings of the 9th ACIS   International Conference on Software Engineering, Artificial Intelligence,   Networking, and Parallel/Distributed Computing & 2008 \\ \hline
58 & Web service test case generation   based on decision table (short paper) & Noikajana and Suwannasart & Proceedings of the 8th   International Conference on Quality Software & 2008 \\ \hline
59 & A novel approach to generate   test cases from UML activity diagrams & Kundu and Samanta & Journal of Object Technology & 2009 \\ \hline
60 & Automatic generation of test   specifications for coverage of system state transitions & Sarma and Mall & Information and Software   Technology & 2009 \\ \hline
61 & Model-based test cases synthesis   using UML interaction diagrams & Nayak and Samanta & ACM SIGSOFT Software Engineering   Notes & 2009 \\ \hline
62 & Test case generation from UML   subactivity and activity diagram & Fan et al. & Proceedings of the 2nd   International Symposium on Electronic Commerce and Security & 2009 \\ \hline
63 & Generating test cases for timed   systems from controlled natural language specifications & Schnelte & Proceedings of the 3rd IEEE   International Conference on Secure Software Integration and Reliability   Improvement & 2009 \\ \hline
64 & TSGen: A UML activity   diagram-based test scenario generation tool & Sun et al. & Proceedings of the 2009   International Conference on Computational Science and Engineering & 2009 \\ \hline
65 & Scenario-based test case   generation using event-B models & Malik et al. & Proceedings of the 1st   International Conference on Advances in System Testing and Validation   Lifecycle & 2009 \\ \hline
66 & Tracing requirements in a   model-based testing approach & Abbors et al. & Proceedings of the 1st   International Conference on Advances in System Testing and Validation   Lifecycle & 2009 \\ \hline
67 & Developing comprehensive   acceptance tests from use cases and robustness diagrams & El-Attar and Miller & Requirements Engineering & 2009 \\ \hline
68 & Synthesis of test scenarios   using UML activity diagrams & Nayak and Samanta & Software \& Systems Modeling & 2009 \\ \hline
69 & Slicing-based test case   generation from UML activity diagrams & Samuel and Mall & ACM SIGSOFT Software Engineering   Notes & 2009 \\ \hline
70 & A hybrid genetic algorithm based   test case generation using sequence diagrams & Shirole and Kumar & Proceedings of the 3rd   International Conference on Contemporary Computing & 2010 \\ \hline
71 & Test case generation based on   state and activity models & Swain et al. & Journal of Object Technology & 2010 \\ \hline
72 & Test case generation based on   use case and sequence diagram & Swain et al. & International Journal of   Software Engineering & 2010 \\ \hline
73 & Formal scenario-based   requirements specification and test case generation in healthcare   applications & L{\"o}ffler et al. & Proceedings of the 2010 ICSE   Workshop on Software Engineering in Health Care & 2010 \\ \hline
74 & Efficient test case generation   for validation of UML activity diagrams & Chen et al. & Design Automation for Embedded   Systems & 2010 \\ \hline
75 & Test sequence generation from   communicating UML state charts: An industrial application of symbolic   transition systems & Schwarzl and Peischl & Proceedings of the 10th   International Conference on Quality Software & 2010 \\ \hline
76 & Automated test case generation   from use case: A model based approach & Chen and Li & Proceedings of the 3rd   International Conference on Computer Science and Information Technology & 2010 \\ \hline
77 & Test case generation from   Behavioral  UML models & Swain and Mohapatra & International Journal of   Computer Applications & 2010 \\ \hline
78 & Generating test cases from UML   activity diagrams using the Condition-Classification Tree Method & Kansomkeat et al. & Proceedings of the 2nd   International conference on software technology and engineering & 2010 \\ \hline
79 & A prolific approach for   automated generation of test cases from informal requirements & Chatterjee and Johari & ACM SIGSOFT Software Engineering   Notes & 2010 \\ \hline
80 & A new approach to generating   high quality test cases & Liu and Miao & Proceedings of the 19th IEEE   Asian Test Symposium & 2010 \\ \hline
81 & Requirement-based test case   generation and prioritization & Salem and Hassan & Proceedings of the 2010   International Computer Engineering Conference & 2010 \\ \hline
82 & An approach to automatic   generation of test cases based on use cases in the requirements phase & Kumaran et al. & International Journal on   Computer Science and Engineering & 2011 \\ \hline
83 & Test Case Generation for   Concurrent System using UML Combinational Diagram & Khandai et al. & International Journal of   Computer Science and Information Technologies & 2011 \\ \hline
84 & A proposed test case generation   technique based on activity diagrams & Boghdady et al. & International Journal of   Engineering & 2011 \\ \hline
85 & Model based approach to assist   test case creation, execution, and maintenance for test automation & Gupta and Surve & Proceedings of the 1st   International Workshop on End-to-End Test Script Engineering & 2011 \\ \hline
86 & Generation of improved test   cases from UML state diagram using genetic algorithm & Shirole et al. & Proceedings of the 4th India   Software Engineering Conference & 2011 \\ \hline
87 & A model-based testing using   scenarios and constraints-based modular petri nets & Reza and Kerlin & Proceedings of the 8th   International Conference on Information Technology: New Generations & 2011 \\ \hline
88 & A novel approach of test case   generation for concurrent systems using UML Sequence Diagram & Khandai et al. & Proceedings of the 3rd   International Conference on Electronics Computer Technology & 2011 \\ \hline
89 & Generating model-based test   cases from natural language requirements for space application software & Santiago J{\'u}nior and Vijaykumar & Software Quality Journal & 2011 \\ \hline
90 & Automation of test case   generation from textual use cases & Jiang and Ding & Proceedings of the 4th   International Conference on Interaction Sciences & 2011 \\ \hline
91 & An enhanced test case generation   technique based on activity diagrams & Boghdady et al. & Proceedings of the 2011   International Conference on Computer Engineering \& Systems & 2011 \\ \hline
92 & Test case generation approach   for industrial automation systems & Hametner et al. & Proceedings of the 5th   International Conference on Automation, Robotics and Applications & 2011 \\ \hline
93 & A" Vibration" method   for automatically generating test cases based on formal specifications & Liu and Nakajima & Proceedings of the 18th   Asia-Pacific Software Engineering Conference & 2011 \\ \hline
94 & Automatic test case generation   from functional requirements in NDT & Guti{\'e}rrez et al. & Proceedings of the 12th   International Conference on Web Engineering & 2012 \\ \hline
95 & Test case generation for   concurrent object-oriented systems using combinational UML models & Dalai et al. & International Journal of   Advanced Computer Science and Applications & 2012 \\ \hline
96 & Litmus: Generation of test cases   from functional requirements in natural language & Dwarakanath and Sengupta & Proceedings of the 17th   International Conference on Applications of Natural Language Processing and   Information Systems & 2012 \\ \hline
97 & An enhanced technique for   generating hybrid coverage test cases using activity diagrams & Boghdady et al. & Proceedings of the 8th   International Conference on Informatics and Systems & 2012 \\ \hline
98 & Functional and user interface   model for generating test cases & Bin and Anbao & Proceedings of the 11th   International Conference on Computer and Information Science & 2012 \\ \hline
99 & Generation test case from UML   activity diagram based on AC grammar & Pechtanun and Kansomkeat & Proceedings of the 2012   International Conference on Computer \& Information Science & 2012 \\ \hline
100 & Practical Application of UML   activity diagrams for the generation of test cases & Fernandez-Sanz and Misra & Proceedings of the Romanian   Academy, Series A & 2012 \\ \hline
101 & Test generation from state based   use case models & Nogueira et al. & Formal Aspects of Computing & 2012 \\ \hline
102 & Automatic generation of   multi-testing types test cases using requirements-based testing & Boghdady et al. & Proceedings of the 7th   International Conference on Computer Engineering \& Systems & 2012 \\ \hline
103 & Test case generation based on   activity diagram for mobile application & Chouhan et al. & International Journal of   Computer Applications & 2012 \\ \hline
104 & Automatic test cases generation   from UML activity diagrams using graph transformation & Hettab et al. & Proceedings of the 6th   International Conference on Information Technology & 2013 \\ \hline
105 & Test case generation from   natural language requirements based on SCR specifications & Carvalho et al. & Proceedings of the 28th Annual   ACM Symposium on Applied Computing & 2013 \\ \hline
106 & Extenics-based test case   generation for UML activity diagram & Li et al. & Procedia Computer Science & 2013 \\ \hline
107 & A model based approach to test   case generation for testing the navigation behavior of dynamic web   applications & Bansal and Sabharwal & Proceedings of the 6th   International Conference on Contemporary Computing & 2013 \\ \hline
108 & Automatic generation of test   cases for critical systems based on MC/DC criteria & Almeida et al. & Proceedings of the IEEE/AIAA   32nd Digital Avionics Systems Conference & 2013 \\ \hline
109 & An approach to generate safety   validation test cases from UML activity diagram & Tiwari and Gupta & Proceedings of the 20th   Asia-Pacific Software Engineering Conference & 2013 \\ \hline
110 & Generation of test cases from   software requirements using natural language processing & Verma and Beg & Proceedings of the 6th   International Conference on Emerging Trends in Engineering and Technology & 2013 \\ \hline
111 & An automated approach of test   case generation for concurrent systems from requirements descriptions & Sarmiento et al. & Proceedings of the 16th   International Conference on Enterprise Information Systems & 2014 \\ \hline
112 & NAT2TESTSCR: Test case   generation from natural language requirements based on SCR specifications & Carvalho et al. & Science of Computer Programming & 2014 \\ \hline
113 & A systematic approach to   automatically derive test cases from use cases specified in restricted   natural languages & Zhang et al. & Proceedings of the 2014   International Conference on System Analysis and Modeling & 2014 \\ \hline
114 & Model-based testing from   controlled natural language requirements & Carvalho et al. & Formal Techniques for   Safety-Critical Systems & 2014 \\ \hline
115 & A novel approach for test case   generation from UML activity diagram & Jena et al. & Proceedings of the 2014   International Conference on Issues and Challenges in Intelligent Computing   Techniques & 2014 \\ \hline
116 & An approach to generate test   cases from sequence diagram & Dhineshkumar and Galeebathullah & Proceedings of the 2014   International Conference on Intelligent Computing Applications & 2014 \\ \hline
117 & Automated generation of test   cases from logical specification of software requirements & Sharma and Biswas & Proceedings of the 9th   International Conference on Evaluation of Novel Approaches to Software   Engineering & 2014 \\ \hline
118 & Test case generation using UML   state diagram and OCL expression & AzaharuddinAli et al. & International Journal of   Computer Applications & 2014 \\ \hline
119 & Generation of test cases using   activity diagram & Swain et al. & International Journal of   Computer Science and Informatics & 2014 \\ \hline
120 & Keeping requirements and test   cases consistent: Towards an ontology-based approach & Feldmann et al. & Proceedings of the 12th IEEE   International Conference on Industrial Informatics & 2014 \\ \hline
121 & Generation of test cases from   sliced sequence diagram & Kaur and Singh & International Journal of   Computer Applications & 2014 \\ \hline
122 & Towards the automated generation   of abstract test cases from requirements models & Granda et al. & Proceedings of the IEEE 1st   International Workshop on Requirements Engineering and Testing & 2014 \\ \hline
123 & C\&L: Generating model based   test cases from natural language requirements descriptions & Sarmiento et al. & Proceedings of the IEEE 1st   International Workshop on Requirements Engineering and Testing & 2014 \\ \hline
124 & Abstract test case generation   for behavioural testing of software product lines & Devroey et al. & Proceedings of the 18th   International Software Product Line Conference: Companion Volume for   Workshops, Demonstrations and Tools-Volume 2 & 2014 \\ \hline
125 & Integration of requirements   engineering and test-case generation via OSLC & Aichernig et al. & Proceedings of the 14th   International Conference on Quality Software & 2014 \\ \hline
126 & Cost-effective functional   testing of reactive software & Venkatesh et al. & Proceedings of the 10th   International Conference on Evaluation of Novel Approaches to Software   Engineering & 2015 \\ \hline
127 & On generating test cases from   EDT specifications & Venkatesh et al. & International Conference on   Evaluation of Novel Approaches to Software Engineering & 2015 \\ \hline
128 & Functional test cases generation   based on automated generated use case diagram & Singh and Sharma & International Journal of   Innovative Research in Advanced Engineering & 2015 \\ \hline
129 & A model-driven approach for   functional test case generation & Guti{\'e}rrez et al. & Journal of Systems and Software & 2015 \\ \hline
130 & Test case generation and   optimization using UML models and genetic algorithm & Khurana and Chillar & Procedia Computer Science & 2015 \\ \hline
131 & A graph transformation approach   for automatic test cases generation from UML activity diagrams & Hettab et al. & Proceedings of the 8th   International C* Conference on Computer Science \& Software Engineering & 2015 \\ \hline
132 & Automatic generation of system   test cases from use case specifications & Wang et al. & Proceedings of the 2015   International Symposium on Software Testing and Analysis & 2015 \\ \hline
133 & An approach of generating test   requirements for agile software development & Tiwari and Gupta & Proceedings of the 8th India   Software Engineering Conference & 2015 \\ \hline
134 & Natural language generation   approach for automated generation of test cases from logical specification of   requirements & Sharma and Biswas & Proceedings of the 2014   International Conference on Evaluation of Novel Approaches to Software   Engineering & 2015 \\ \hline
135 & Scalable incremental test-case   generation from large behavior models & Aichernig et al. & Proceedings of the 2015   International Conference on Tests and Proofs & 2015 \\ \hline
136 & NAT2TEST tool: From natural   language requirements to test cases based on CSP & Carvalho et al. & Software Engineering and Formal   Methods & 2015 \\ \hline
137 & An automated test generation   technique for software quality assurance & Xu et al. & IEEE Transactions on Reliability & 2015 \\ \hline
138 & Model-based approach for   automated test case generation from visual requirement specifications & Singi et al. & Proceedings of the IEEE 8th   International Conference on Software Testing, Verification and Validation   Workshops & 2015 \\ \hline
139 & Automation of test case   generation from behavior tree requirements models & Lindsay et al. & Proceedings of the 24th   Australasian Software Engineering Conference & 2015 \\ \hline
140 & Towards the generation of tests   in the test description language from use case map models & Boulet et al. & Proceedings of the 17th   International SDL Forum on SDL 2015: Model-Driven Engineering for Smart   Cities & 2015 \\ \hline
141 & Waterfall and agile   requirements-based model for automated test cases generation & Elghondakly et al. & Proceedings of the IEEE 7th   International Conference on Intelligent Computing and Information Systems & 2015 \\ \hline
142 & Automatic generation of test   cases from activity diagrams for UML based testing (UBT) & Oluwagbemi and Asmuni & Jurnal Teknologi (Sciences   \& Engineering) & 2015 \\ \hline
143 & ACT (Abstract to Concrete Tests)   - A tool for generating Concrete test cases from formal specification of web   applications & Bubna and Chakrabarti & Proceedings of the 2nd Modelling   Symposium & 2016 \\ \hline
144 & A novel technique for generation   and optimization of test cases using use case, sequence, activity diagram and   genetic algorithm & Khurana et al. & Journal of Software & 2016 \\ \hline
145 & An optimized approach for   automated test case generation and validation for UML diagrams & Elghondakly et al. & Asian J Inf Technol & 2016 \\ \hline
146 & A transformation-based approach   to testing concurrent programs using UML activity diagrams & Sun et al. & Software: Practice and   Experience & 2016 \\ \hline
147 & Automatic test generation from   UML sequence diagrams for Android mobiles & Anbunathan and Basu & International Journal of Applied   Engineering Research & 2016 \\ \hline
148 & Automatic generation of   specification-based test cases by applying genetic algorithms in   reinforcement learning & Sato and Sugihara & International Workshop on   Structured Object-Oriented Formal Language and Method & 2016 \\ \hline
149 & An optimal solution for test   case generation using ROBDD graph and PSO algorithm & Kalaee and Rafe & Quality and Reliability   Engineering International & 2016 \\ \hline
150 & Test scenario generation from   natural language requirements descriptions based on Petri-Nets & Sarmiento et al. & Electronic Notes in Theoretical   Computer Science & 2016 \\ \hline
151 & Generating test cases from UML   activity diagram based on business flow constraints & Thanakorncharuwit et al. & Proceedings of the 5th   International Conference on Network, Communication and Computing & 2016 \\ \hline
152 & Syntactic rules of extracting   test cases from software requirements & Masuda et al. & Proceedings of the 8th   International Conference on Information Management and Engineering & 2016 \\ \hline
153 & Test case generation from   natural language requirements using CPN simulation & Silva et al. & Proceedings of the 2015   Brazilian Symposium on Formal Methods & 2016 \\ \hline
154 & Automatic generation of test   cases and test purposes from natural language & Nogueira et al. & Formal Methods: Foundations and   Applications & 2016 \\ \hline
155 & Automatic test case generation   from requirements for industrial cyber-physical systems & Sinha et al. & at-Automatisierungstechnik & 2016 \\ \hline
156 & Utilizing business process   models to generate software test cases & Khader and Yousef & International Journal of   Computer Science Issues & 2016 \\ \hline
157 & TestMEReq: generating abstract   tests for requirements validation & Moketar et al. & Proceedings of the 3rd   International Workshop on Software Engineering Research and Industrial   Practice & 2016 \\ \hline
158 & Automatic generation of TestNG   tests cases from UML sequence diagrams in Scrum process & Elallaoui et al. & Proceedings of the 4th IEEE   International Colloquium on Information Science and Technology & 2016 \\ \hline
159 & A search-based approach for test   suite generation from extended finite state machines & Rao et al. & Proceedings of the 2016   International Conference on Progress in Informatics and Computing & 2016 \\ \hline
160 & Application of inference rules   to a software requirements ontology to generate software test cases & Tarasov et al. & Proceedings of the 13th   International Workshop on Experiences and Directions and 5th International   Workshop on Reasoner Evaluation & 2017 \\ \hline
161 & Automated user story driven   approach for web based functional testing & Masud et al. & International Journal of   Computer and Information Sciences & 2017 \\ \hline
162 & From statecharts into model   checking: A hierarchy-based translation and specification patterns properties   to generate test cases & de Santiago J{\'u}nior  and da   Silva & Proceedings of the 2nd Brazilian   Symposium on Systematic and Automated Software Testing & 2017 \\ \hline
163 & Generating test cases and   procedures from use cases in dynamic software product lines & Ara{\'u}jo et al. & Proceedings of the Symposium on   Applied Computing & 2017 \\ \hline
164 & Automated test case generation   from UML activity diagram and sequence diagram using depth first search   algorithm & Meiliana et al. & Procedia Computer Science & 2017 \\ \hline
165 & Constructing test cases using   natural language processing & Ansari et al. & Proceedings of the 3rd   International Conference on Advances in Electrical, Electronics, Information,   Communication and Bio-Informatics & 2017 \\ \hline
166 & Transforming workflow models   into automated end-to-end acceptance test cases & Boucher and Mussbacher & Proceedings of the IEEE/ACM 9th   International Workshop on Modelling in Software Engineering & 2017 \\ \hline
167 & Automated test case generation   from high-level logic requirements using model   transformation techniques & Olajubu et al. & Proceedings of the 9th Computer   Science and Electronic Engineering & 2017 \\ \hline
168 & Towards the generation of   end-to-end web test scripts from requirements specifications & Clerissi et al. & Proceedings of the IEEE 25th   International Requirements Engineering Conference Workshops & 2017 \\ \hline
169 & Reqcap: Hierarchical   requirements modeling and test generation for industrial control systems & Almohammad et al. & Proceedings of the IEEE 25th   International Requirements Engineering Conference Workshops & 2017 \\ \hline
170 & An automated test case   generation approach based on activity diagrams of SysML & Yin et al. & International Journal of   Performability Engineering & 2017 \\ \hline
171 & Automatic test case generation   through collaboration diagram: a case study & Kaur and Vig & International Journal of System   Assurance Engineering and Management & 2017 \\ \hline
172 & Requirement-based   testing-extracting logical test cases from requirement documents & Sneed & Software Quality: Methods and   Tools for Better Software and Systems & 2017 \\ \hline
173 & An automated functional testing   framework for context-aware applications & Mirza and Khan & IEEE Access & 2018 \\ \hline
174 & Named entity recognition for   automated test case generation & Mahalakshmi et al. & The International Arab Journal   of Information Technolog & 2018 \\ \hline
175 & Automatic test case generation   method for large scale communication node software & Kikuma et al. & Proceedings of the 2018   International Conference on Emerging Internetworking, Data \& Web   Technologies & 2018 \\ \hline
176 & Automated generation of   requirements-based test cases for an adaptive cruise control system & Aniculaesei et al. & 2018 IEEE Workshop on   Validation, Analysis and Evolution of Software Tests & 2018 \\ \hline
177 & Automatic generation of test   cases from  UML   models & P{\'e}rez  and Mar{\'\i}n & CLEI Electronic Journal & 2018 \\ \hline
178 & Automatic test transition paths   generation approach from EFSM using state tree & Chen et al. & Proceedings of the 2018 IEEE   International Conference on Software Quality, Reliability and Security   Companion & 2018 \\ \hline
179 & From natural language   requirements to passive test cases using guarded assertions & Danie et al. & Proceedings of the 2018 IEEE   International Conference on Software Quality, Reliability and Security & 2018 \\ \hline
180 & Automatic test cases generation   from business process models & Yazdani et al. & Requirements Engineering & 2018 \\ \hline
181 & A transformation-based method   for test case automatic generation from use cases & Hue et al. & Proceedings of the 10th   International Conference on Knowledge and Systems Engineering & 2018 \\ \hline
182 & Test scenarios generation using   UML sequence diagram & Panigrahi et al. & Proceedings of the 2018   International Conference on Information Technology & 2018 \\ \hline
183 & From modeling to test case   generation in the industrial embedded system domain & Hussain et al. & Proceedings of the 2018   International Conferences on Software Technologies: Applications and   Foundations & 2018 \\ \hline
184 & Automated generation of   requirements-based test cases for an automotive function using the SCADE   toolchain & Aniculaesei et al. & Proceedings of the 11th   International Conference on Adaptive and Self-Adaptive Systems and   Applications & 2019 \\ \hline
185 & An efficient method to generate   test cases from UML-use case diagram & Khalifa and Jamil & International Journal of   Engineering Research and Technology & 2019 \\ \hline
186 & From requirements to automated   acceptance tests of interactive apps: An integrated model-based testing   approach & Maciel et al. & Proceedings of the 14th   International Conference on Evaluation of Novel Approaches to Software   Engineering & 2019 \\ \hline
187 & CPN simulation-based test case   generation from controlled natural-language requirements & Silva et al. & Science of Computer Programming & 2019 \\ \hline
188 & A memorization approach for test   case generation in concurrent UML activity diagram & Kamonsantiroj et al. & Proceedings of the 2019 2nd   International Conference on Geoinformatics and Data Analysis & 2019 \\ \hline
189 & Ontology based test case   generation for black box testing & Ul Haq and Qamar & Proceedings of the 2019 8th   International Conference on Educational and Information Technology & 2019 \\ \hline
190 & From use case maps to executable   test procedures: a scenario-based approach & Kesserwan et al. & Software \& Systems Modeling & 2019 \\ \hline
191 & Using OWL ontologies as a   domain-specific language for capturing requirements for formal analysis and   test case generation & Crapo and Moitra & Proceedings of the IEEE 13th   International Conference on Semantic Computing & 2019 \\ \hline
192 & MCCFG: an MOF-based multiple   condition control flow graph for automatic test case generation & Son et al. & Cluster Computing & 2019 \\ \hline
193 & Test case generation using   unified modeling language & Shah et al. & Proceedings of the International   Conference on Computer and Information Sciences & 2019 \\ \hline
194 & An automated approach to   generate test cases from use case description model & Alrawashed et al. & Computer Modeling in Engineering   \& Sciences & 2019 \\ \hline
195 & Specification-based test case   generation with genetic algorithm & Wang et al. & Proceedings of the 2019 IEEE   Congress on Evolutionary Computation & 2019 \\ \hline
196 & Automating requirements analysis   and test case generation & Moitra et al. & Requirements Engineering & 2019 \\ \hline
197 & Towards transforming user   requirements to test cases using MDE and NLP & Allala et al. & Proceedings of the IEEE 43rd   Annual Computer Software and Applications Conference & 2019 \\ \hline
198 & Test case generation, selection   and coverage from natural language & Nogueira et al. & Science of Computer Programming & 2019 \\ \hline
199 & Using the SCADE toolchain to   generate requirements-based test cases for an adaptive cruise control system & Aniculaesei et al. & Proceedings of the ACM/IEEE 22nd   International Conference on Model Driven Engineering Languages and Systems   Companion & 2019 \\ \hline
200 & USLTG: Test   Case Automatic Generation by Transforming Use Cases & Hue et al. & International Journal of   Software Engineering and Knowledge Engineering & 2019 \\ \hline
201 & Requirements-based automated   test generation for safety critical software & Li et al. & Proceedings of the 38th Digital   Avionics Systems Conference & 2019 \\ \hline
202 & Generating test sequences from   UML use-case diagrams & Hamza and Hammad & Proceedings of the 2019   International Conference on Innovation and Intelligence for Informatics,   Computing, and Technologies & 2019 \\ \hline
203 & Enhancing software testing with   ontology engineering approach & Charoenreh and Intana & Proceedings of the 23rd   International Computer Science and Engineering Conference & 2019 \\ \hline
204 & Impact analysis framework of   test cases based on changes of use case based requirements & Intana and Sriraksa & Proceedings of the 23rd   International Computer Science and Engineering Conference & 2019 \\ \hline
205 & Knowledge graph based automated   generation of test cases in software engineering & Nayak et al. & Proceedings of the 7th ACM IKDD   CoDS and 25th COMAD & 2020 \\ \hline
206 & Test case generation algorithms   and tools for specifications in natural language & Aoyama et al. & Proceedings of the 2020 IEEE   International Conference on Consumer Electronics & 2020 \\ \hline
207 & From requirements to automated   acceptance tests with the RSL language & Paiva et al. & Proceedings of the 2019   International Conference on Evaluation of Novel Approaches to Software   Engineering & 2020 \\ \hline
208 & Automatic generation and   optimization of test case using hybrid cuckoo search and bee colony algorithm & Lakshminarayana and SureshKumar & Journal of Intelligent Systems & 2020 \\ \hline
209 & Model-based testing from model   driven architecture: A novel approach for automatic test cases generation & Essebaa et al. & Proceedings of the 2019   International Conference on Smart Information \& Communication   Technologies & 2020 \\ \hline
210 & Preliminary experiences in   requirements-based security testing & Miranda et al. & Proceedings of the 2020   International Conference on the Quality of Information and Communications   Technology & 2020 \\ \hline
211 & Scaling test case generation for   expressive decision tables & Agrawa et al. & Proceedings of the IEEE 13th   International Conference on Software Testing, Validation and Verification & 2020 \\ \hline
212 & SPECMATE: Automated creation of   test cases from acceptance criteria & Fischbach et al. & Proceedings of the IEEE 13th   International Conference on Software Testing, Validation and Verification & 2020 \\ \hline
213 & SYNTest: Prototype of syntax   test case generation tool & Intana et al. & Proceedings of the 5th   International Conference on Information Technology & 2020 \\ \hline
214 & Automatic test case generation   from formal requirement model for avionics software & Wang et al. & Proceedings of the 6th   International Symposium on System and Software Reliability & 2020 \\ \hline
215 & Test generation tool for   modified condition/decision coverage: model based Testing & El qortobi et al. & Proceedings of the 13th   International Conference on Intelligent Systems: Theories and Applications & 2020 \\ \hline
216 & Specification-based test case   generation with constrained genetic programming & Sato & Proceedings of the 20th   International Conference on Software Quality, Reliability and Security   Companion & 2020 \\ \hline
217 & NLP-based requirements   formalization for automatic test case generation & Gr{\"o}pler et al. & Proceedings of the 29th   International Workshop on Concurrency, Specification and Programming & 2021 \\ \hline
218 & Towards a test case generation   tool based on functional requirements & Fernandes and Maciel & Proceedings of the 19th   Brazilian Symposium on Software Quality & 2021 \\ \hline
219 & Test case generation from   UML-diagrams using genetic algorithm & Sahoo et al. & Computers, Materials and   Continua & 2021 \\ \hline
220 & Towards a model-driven testing   framework for GUI test cases generation from user stories & Granda et al. & Proceedings of the 16th   International Conference on Evaluation of Novel Approaches to Software   Engineering & 2021 \\ \hline
221 & Analyzing UML use   cases to generate test sequences & Hamza and Hammad & International Journal of   Computing and Digital Systems & 2021 \\ \hline
222 & Executable test case generation   from specifications written in natural language and test execution   environment & Aoyama et al. & Proceedings of the IEEE 18th   Annual Consumer Communications \& Networking Conference & 2021 \\ \hline
223 & An ontology-based approach to   automated test case generation & Banerjee et al. & SN Computer Science & 2021 \\ \hline
224 & Accuracy improvement by training   data selection in automatic test cases generation method & Ueda and Tsukada & Proceedings of the 9th   International Conference on Information and Education Technology & 2021 \\ \hline
225 & Model-based test case generation   from UML sequence diagrams using extended finite state machines & Rocha et al. & Software Quality Journal & 2021 \\ \hline
226 & A method of test case generation   based on VRM model & Zhipeng et al. & Proceedings of the IEEE 6th   International Conference on Computer and Communication Systems & 2021 \\ \hline
227 & Syntax-tree similarity for   test-case derivability in software requirements & Masuda et al. & Proceedings of the 2021 IEEE   International Conference on Software Testing, Verification and Validation   Workshops & 2021 \\ \hline
228 & Coverage-based grammar-guided   genetic programming generation of test suites & Ibias et al. & Proceedings of the 2021 IEEE   Congress on Evolutionary Computation & 2021 \\ \hline
229 & An improved approach for   generating test cases during model-based testing using tree traversal   algorithm & Achimugu et al. & Journal of Software Engineering   and Applications & 2021 \\ \hline
230 & Automated test cases generation   from requirements specification & Lafi et al. & Proceedings of the 2021   International Conference on Information Technology & 2021 \\ \hline
231 & Generating test scenarios using   SysML activity diagram & Yang et al. & Proceedings of the 8th   International Conference on Dependable Systems and Their Applications & 2021 \\ \hline
232 & Generating test cases from   requirements: A case study in railway control system domain & Zheng et al. & Proceedings of the 2021   International Symposium on Theoretical Aspects of Software Engineering & 2021 \\ \hline
233 & Automatic generation of test   cases from formal specifications using mutation testing & Cajica et al. & Proceedings of the 18th   International Conference on Electrical Engineering, Computing Science and   Automatic Control & 2021 \\ \hline
234 & Model-based test case generation   approach for mobile applications load testing using OCL enhanced activity   diagrams & Ali et al. & Proceedings of the 10th   International Conference on Intelligent Computing and Information Systems & 2021 \\ \hline
235 & A tool to support vibration   testing method for automatic test case generation and test result analysis & Saiki et al. & Proceedings of the IEEE 21st   International Conference on Software Quality, Reliability and Security & 2021 \\ \hline
236 & Forward-propagation approach for   generating feasible and minimum test case suites from cause-effect graph   specifications & Krupalija et al. & IEEE Access & 2022 \\ \hline
237 & Application of the equivalent   partitioning method in testing for automatic test case generation on the   Digi-OTA system & Amalia et al. & Jurnal Media Informatika   Budidarma & 2022 \\ \hline
238 & Automatic generation of   acceptance test cases from use case specifications: An NLP-based approach & Wang et al. & IEEE Transactions on Software   Engineering & 2022 \\ \hline
239 & Automatic test case and test   oracle generation based on functional scenarios in formal specifications for   conformance testing & Liu and Nakajima & IEEE Transactions on Software   Engineering & 2022 \\ \hline
240 & Transition coverage based test   case generation from state chart diagram & Pradhan et al. & Journal of King Saud   University-Computer and Information Sciences & 2022 \\ \hline
241 & Automating test case generation   for android applications using model-based testing & Khan et al. & EMITTER International Journal of   Engineering Technology & 2022 \\ \hline
242 & Prioritized variable-length test   cases generation for finite state machines & Rechtberger et al. & Proceedings of the 2022 IEEE   International Conference on Software Testing, Verification and Validation   Workshops & 2022 \\ \hline
243 & Requirements-driven model   checking and test generation for comprehensive verification & Bhatt et al. & Proceedings of the 14th   International Symposium NASA Formal Methods & 2022 \\ \hline
244 & Model-based automated testing   method for airborne system & Zhu et al. & Proceedings of the 5th   International Conference on Aerospace System Science and Engineering & 2022 \\ \hline
245 & MC/DC test   case automatic generation for safety-critical systems & Wang et al. & Proceedings of the IEEE 22nd   International Conference on Software Quality, Reliability and Security & 2022 \\ \hline
246 & An improved test case generation   method based on test requirements for testing software component & Chen et al. & Proceedings of the IEEE 22nd   International Conference on Software Quality, Reliability, and Security   Companion & 2022 \\ \hline
247 & Generating abstract test cases   from user requirements using MDSE and NLP & Allala et al. & Proceedings of the IEEE 22nd   International Conference on Software Quality, Reliability and Security & 2022 \\ \hline
248 & Model-checking-based automated   test case generation for Z formal specification of an urban railway   interlocking system & Kadakolmath and Ramu & Proceedings of the 4th   International Conference on Emerging Research in Electronics, Computer   Science and Technology & 2022 \\ \hline
249 & Automatic creation of acceptance   tests by extracting conditionals from requirements: NLP approach and case   study & Fischbach et al. & Journal of Systems and Software & 2023 \\ \hline
250 & Automated test case generation   using T5 and GPT-3 & Mathur et al. & Proceedings of the 9th   International Conference on Advanced Computing and Communication Systems & 2023 \\ \hline
251 & Generating robot framework test   scripts from user stories and scenarios for web application testing & Waitchasarn et al. & Proceedings of the 8th   International Conference on Computer and Communication Systems & 2023 \\ \hline
252 & Research on software test case   generation based on SysML & Chen et al. & Proceedings of the 2023   International Symposium on Intelligent Robotics and Systems & 2023 \\ \hline
253 & Decision tree based test case   generation using story board and natural language processing & Gupta et al. & Proceedings of the 2023   International Conference on Advances in Computing and Data Sciences & 2023 \\ \hline
254 & Transforming software   requirements into test cases via model transformation & Kesserwan et al. & International Journal of   Software Engineering \& Applications & 2023 \\ \hline
255 & An approach of test case   generation with software requirement ontology & Intana et al. & International Journal of   Advanced Computer Science and Applications & 2023 \\ \hline
256 & Granular traceability between   requirements and test cases for safety-critical software systems & Elqortobi et al. & Proceedings of the 19th   International Conference on Mobile Web and Intelligent Information Systems & 2023 \\ \hline
257 & Automatic test sequence   generation and functional coverage measurement from UML sequence diagrams & Ekici and Tuglular & International Journal of   Information System Modeling and Design & 2023 \\ \hline
258 & Test case generation and history   data analysis during optimization in regression testing: An NLP study & Gupta and Mahapatra & Cogent Engineering & 2023 \\ \hline
259 & Generating automated test case   from sequence diagram using Pre-order Traversal & Yonathan et al. & Procedia Computer Science & 2024 \\ \hline
260 & Test case information extraction   from requirements specifications using NLP-based unified boilerplate approach & Lim et al. & Journal of Systems and Software & 2024 \\ \hline
261 & A method for selecting training   data using Doc2Vec for automatic test Cases generation & Fujita and Ueda & Proceedings of the 2024 IEEE   International Conference on Consumer Electronics & 2024 \\ \hline
262 & A case study: To discuss the   construction of the finite state machine and markov-chain OP models for test   case generation & Shaheen et al. & Proceedings of the 5th   International Conference on Advancements in Computational Sciences & 2024 \\ \hline
263 & Generating test scenarios from   NL requirements using retrieval-augmented LLMs: An industrial study & Arora et al. & Proceedings of the IEEE 32nd   International Requirements Engineering Conference & 2024 \\ \hline
264 & Automatic test cases generation   with selection of training data for various system specifications & Ueda et al. & Proceedings of the 6th   International Conference on Computer Communication and the Internet & 2024 \\ \hline
265 & User story based automated test   case generation using NLP & Chinnaswamy et al. & Proceedings of the 2024   International Conference on Computational Intelligence in Data Science & 2024 \\ \hline
266 & Automated test case generation for satellite FRD using NLP and   large language model & Shakthi et al. & Proceedings of the 4th   International Conference on Electrical, Computer, Communications and   Mechatronics Engineering & 2024 \\ \hline
267 & Leveraging pre-trained large language models (LLMs) for   on-premises comprehensive automated test case generation: An empirical study & Yin et al. & Proceedings of the 9th   International Conference on Intelligent Informatics and Biomedical Sciences & 2024 \\ \hline
\end{longtable}
\end{scriptsize}

\end{document}